\documentclass[english,10pt,osajnl2,twocolumn,showpacs]{revtex4}

\usepackage{amssymb,amsmath}
\usepackage[latin1]{inputenc}
\usepackage{txfonts}
\usepackage[dvipsnames,usenames]{color}
\usepackage{graphicx}
\graphicspath{{./}{./figs/}}
\usepackage{ifthen}

\usepackage[pdfpagemode=None,pdfstartview=FitB,pdfpagemode=None]{hyperref}
\hypersetup{
  breaklinks=true,
  colorlinks=true,
  linkcolor=MidnightBlue,
  citecolor=MidnightBlue,
  filecolor=MidnightBlue,
  menucolor=MidnightBlue,
  pagecolor=MidnightBlue,
  urlcolor=MidnightBlue,
}

\makeatletter

\makeatletter
\newcommand{\Symbol}[2]{
  \begingroup \escapechar\m@ne\xdef\my@tempa{\string#1}\endgroup
  \expandafter\@ifundefined{\my@tempa}{%
    \def\my@tempb{[}
    \expandafter\edef\csname\my@tempa\endcsname{%
      \noexpand\@ifnextchar\my@tempb%
          {\csname\my@tempa @b\endcsname}%
          {\csname\my@tempa @a\endcsname}%
    }
    \expandafter\def\csname\my@tempa @a\endcsname{%
      #2%
    }
    \expandafter\def\csname\my@tempa @b\endcsname[##1]{%
      {{#2}_{##1}}%
    }
  }{\@latex@error{\noexpand#1is already defined}\@ehc}%
}
\newcommand{\SymbolWithLabel}[3]{%
  \begingroup \escapechar\m@ne\xdef\my@tempa{\string#1}\endgroup
  \expandafter\@ifundefined{\my@tempa}{%
    \expandafter\edef\csname\my@tempa\endcsname{%
      \noexpand\@ifnextchar_%
          {\csname\my@tempa @b\endcsname}%
          {\csname\my@tempa @a\endcsname}%
    }
    \expandafter\def\csname\my@tempa @a\endcsname{{#2}_{#3}}
    \expandafter\def\csname\my@tempa @b\endcsname{{#2}^{#3}}
  }{\@latex@error{\noexpand#1is already defined}\@ehc}%
}
\newcommand{\SymbolWithLabelAndOptionalArgs}[3]{%
  \begingroup \escapechar\m@ne\xdef\my@tempa{\string#1}\endgroup
  \expandafter\@ifundefined{\my@tempa}{%
    \def\my@tempb{[} 
    \expandafter\edef\csname\my@tempa\endcsname{%
      \noexpand\@ifnextchar\my@tempb%
          {\csname\my@tempa @b\endcsname}%
          {\csname\my@tempa @a\endcsname}%
    }
    \expandafter\def\csname\my@tempa @a\endcsname{%
      {#2}_{#3}%
    }
    \expandafter\def\csname\my@tempa @b\endcsname[##1]{%
      {{#2}^{#3}_{##1}}%
    }
  }{\@latex@error{\noexpand#1is already defined}\@ehc}%
}
\makeatother

 \newcommand{\RM}[1]{\mathrm{#1}}         
 \newcommand{\BS}[1]{{\boldsymbol{#1}}}   
 \newcommand{\V}[1]{\boldsymbol{#1}}      
 \newcommand{\M}[1]{\mathbf{#1}}          
 \newcommand{\T}{^\mathrm{T}}             
 \newcommand{\mT}{^{-\mathrm{T}}}         
 \newcommand{\D}[1]{{\mathrm{d}#1}}       
 \newcommand{\Var}{\operatorname{Var}}    
 \newcommand{\diag}{\operatorname{diag}}  
 \newcommand{\ScaleAs}{\mathcal{O}}       

\newcommand{\norm}[1]{\Vert #1\Vert}
\newcommand{\Norm}[1]{\left\Vert #1\right\Vert}
\newcommand{\abs}[1]{\vert #1\vert}

\newcommand{\avg}[1]{\langle #1\rangle}

\newcommand{\bydef}{\stackrel{\mathrm{def}}{=}}

\newcommand{\Matrix}[2]{\left(\begin{array}{#1}#2\end{array}\right)}

\newcommand{\Cerror}{\M{C}_{\V{n}}}      
\newcommand{\Cprior}{\M{C}_{\V{w}}}      

\newcommand{\Ndata}{M}                     
\newcommand{\Nparam}{N}                    
\newcommand{\Ndif}{N_\mathrm{dif}}         
\newcommand{\Nint}{N_{\mathrm{int}}}       
\SymbolWithLabel{\Niter}{N}{\mathrm{iter}} 
\SymbolWithLabel{\Nops}{N}{\mathrm{ops}}   
\newcommand{\Noverhead}{N_\mathrm{overhead}}

\newcommand{\DoubleSumNint}[2]{\!\!\!\!\sum_{1\le #1<#2\le \Nint}\!\!\!\!}

\newcommand{\argmin}{\mathop{\mathrm{arg\,min}}\limits}

\newcommand{\Eq}[1]{Eq.~(\ref{#1})}
\newcommand{\Fig}[1]{Fig.~\ref{#1}}

\newcommand{\NormalLaw}{\mathcal{N}}     

\usepackage{xspace}
\newcommand{\cf}{\emph{cf.}\xspace}      
\newcommand{\eg}{\emph{e.g.}\xspace}     
\newcommand{\ie}{\emph{i.e.}\xspace}     
\newcommand{\etal}{\emph{et al.}\xspace} 
\newcommand{\etc}{\emph{etc.}\xspace}    

\newcommand{\FRIM}{FRiM\xspace}

\newcommand{\TmpI}{}
\newcommand{\TmpII}{}
\newcommand{\TmpIII}{}
\newcommand{\TmpIV}{}
\newcommand{\TmpV}{}

\newcommand{\SqrtTwo}{{\scriptstyle\!\sqrt{2}}}
\newcommand{\OneOverTwo}{{\textstyle\frac{1}{2}}}

\definecolor{BrightGreen}{rgb}{0.8,1.0,0.0}
\definecolor{BrightBlue}{rgb}{0.5,0.9,1.0}
\definecolor{FireBrick}{rgb}{0.70,0.13,0.13}
\definecolor{Yellow}{rgb}{0.95,0.95,0.00}
\definecolor{Black}{rgb}{0.0,0.0,0.0}

\makeatother

\begin{document}

\title{Fast minimum variance wavefront reconstruction for extremely
  large telescopes}

\author{Eric Thi\'ebaut}

\author{Michel Tallon}

\affiliation{%
  Universit\'e de Lyon, F-69000 Lyon, France;
  Universit\'e de Lyon 1, F-69622 Villeurbanne, France;
  Centre de Recherche Astrophysique de Lyon, Observatoire de Lyon,
  9 avenue Charles Andr\'e, F-69561 Saint-Genis Laval cedex, France;
  CNRS, UMR 5574;
  Ecole Normale Sup\'erieure de Lyon, F-69007 Lyon, France.
}

\email{thiebaut@obs.univ-lyon1.fr}


\begin{abstract}
We present a new algorithm, \FRIM (FRactal Iterative Method), aiming at
the reconstruction of the optical wavefront from measurements provided by a
wavefront sensor.  As our application is adaptive optics on extremely large
telescopes, our algorithm was designed with speed and best quality in mind.
The latter is achieved thanks to a regularization which enforces prior
statistics.  To solve the regularized problem, we use the conjugate
gradient method which  takes advantage of the sparsity of the wavefront
sensor model matrix and avoids the storage and inversion of a huge matrix.
The prior covariance matrix is however non-sparse and we derive a fractal
approximation to the Karhunen-Lo\`eve basis thanks to which the
regularization by Kolmogorov statistics can be computed in
$\ScaleAs(\Nparam)$ operations, $\Nparam$ being the number of phase samples
to estimate.  Finally, we propose an effective preconditioning which also
scales as $\ScaleAs(\Nparam)$ and yields the solution in 5--10 conjugate
gradient iterations for any $\Nparam$.  The resulting algorithm is
therefore $\ScaleAs(\Nparam)$.  As an example, for a $128 \times 128$
Shack-Hartmann wavefront sensor, \FRIM appears to be more than 100 times
faster than the classical vector-matrix multiplication method.
\end{abstract}


\ocis{
  010.7350, 
  100.3190, 
  110.1080. 
}

\maketitle

\section{Introduction}

The standard and most used method for adaptive optics (AO) control is based
on a vector-matrix multiply (VMM) of the vector of wavefront sensor
measurements by the so-called control matrix \cite{Roddier_1999}.  This
operation gives an update of the commands to be sent to the deformable
mirrors to adjust the correction of the corrugated incoming wavefronts.
The control matrix is precomputed, generally using modal control
optimization \cite{Gendron_&_Lena_1994}.  The complexity of computing the
control matrix using standard methods scales as $\ScaleAs(\Nparam^3)$,
where $\Nparam$ is the number of unknowns (phase samples or actuator
commands), and applying real time VMM scales as $\ScaleAs(\Nparam^2)$.  This
computational burden can be reasonably handled on current AO systems where
$\Nparam\lesssim10^{3}$.

For future Extremely Large Telescopes (ELT's), the number of
actuators beeing considered is in the range $10^4-10^5$.  This huge increase is the
result of both the larger diameter of the ELTs \cite{LeLouarn_et_al_2000b}
and the emergence of new architectures for the AO systems, using either a
greater density of actuators (Extreme AO) or combining several deformable
mirrors and wavefront sensors (multi-conjugate AO, multi-object AO)
\cite{Hubin_et_al_2005}.  The necessary computational power for real time
control on such systems is currently unattainable when using standard
methods.

More efficient algorithms are thus required and have been developed in
recent years. Poyneer \etal \cite{Poyneer_et_al_2002} have derived an
accurate Fourier transform wavefront reconstructor by solving the boundary
problem in circular apertures. This reconstructor scales as
$\ScaleAs(\Nparam\log\Nparam)$ and is shown to be effective for Extreme AO
\cite{Poyneer_et_al_2008}. MacMartin \cite{MacMartin_2003} studied several
approximate approaches such as a multiple-layer hierarchic reconstruction,
which scales as $\ScaleAs(\Nparam)$.

Although least-squares algorithms give suitable results for single star AO
systems (classical on-axis AO or Extreme AO), minimum variance reconstruction is
required to minimize the effects of the missing data or unseen modes in the
other AO schemes \cite{LeRoux_et_al_2004}. In the context of minimum
variance for multi-conjugate AO, Ellerbroek \cite{Ellerbroek_2002} could
apply sparse matrix techniques (Cholesky factorization) using a sparse
approximation of the turbulence statistics, and introducing as low-rank
adjustments, the nonsparse matrix terms arising from the global tip/tilt
measurement errors associated with laser guide stars. However the interactions
between the layers in their tomographic modeling reduce the efficiency of
the sparse direct decomposition methods \cite{Vogel_2004}.

Iterative methods are also extensively studied in this context. Their main
asset is their ability to iteratively compute the unknowns from the
measurements using direct sparse matrices, and so the storage of a
precomputed inverse full matrix is not necessary. One major problem
with iterative methods is the increase in the number of iterations with the
number of unknowns to estimate \cite{Southwell_1980,
Nocedal_Wright-2006-numerical_optimization, NumericalRecipes}. As an
example, Wild \etal \cite{Wild_et_al_1995} have proposed to use the
closed-loop AO system itself as an iterative processor, but the performance
of the least squares reconstruction depends on the loop frequency of the AO
system, which should be higher than usual.

The most successful iterative methods in AO are now based on preconditioned
conjugate gradients (PCG) \cite{templates}, where some of the previous
approximate reconstruction methods are embedded as preconditioners to ensure a
small number of iterations (see section~\ref{sec:preconditioning}). Gilles
\etal \cite{Gilles_et_al_2002c} have described a multigrid PCG algorithm,
mainly aimed at Extreme AO and scaling as $\ScaleAs(\Nparam\log\Nparam)$.
The multigrid preconditioner is somewhat related to the multiple-layers
hierarchic reconstruction \cite{MacMartin_2003}. This wavefront
reconstruction method has been improved with a faster approximation to the
turbulence statistics, scaling as $\ScaleAs(\Nparam)$ \cite{Gilles_2003b}.
The multigrid PCG algorithm has also been developed for multi-conjugate AO
\cite{Gilles_et_al_2003}. In this case, the structure of the matrix is more
complex and brings some limitations.
%
More recently, a Fourier domain preconditioner was introduced
\cite{Yang_et_al_2006c, Vogel_&_Yang_2006c} in the context of
multi-conjugate AO, with a faster reconstruction than multigrid PCG. In
this case, the preconditioner is related to the Fourier transform wavefront
reconstructor \cite{Poyneer_et_al_2002}. Both multigrid and Fourier domain
preconditioners were examined for the Thirty Meter Telescope project
\cite{Gilles_et_al_2007,Gilles_&_Ellerbroek_2008b}.

In this work, we propose novel methods to address the two critical points
previously seen in iterative methods for wavefront reconstruction:
estimation of the atmospheric phase covariance matrix and preconditioning.

We need a sparse representation of the inverse of the atmospheric phase
covariance matrix to efficiently introduce priors in the minimum variance
estimator. 
Currently, we can choose between a good representation in the Fourier
domain with $\ScaleAs(\Nparam\log\Nparam)$ complexity
\cite{Gilles_et_al_2002c, Yang_et_al_2006c} and a widely used sparse
biharmonic approximation introduced by Ellerbroek \cite{Ellerbroek_2002},
less accurate \cite{Yang_et_al_2006c}, but scaling as $\ScaleAs(\Nparam)$.
With \FRIM, we introduce a so-called ``fractal operator'' as a multiscale
algorithm with $\ScaleAs(\Nparam)$ complexity. This operator, both accurate
and very fast, was inspired by the mid-point method of Lane \etal
\cite{LaneEtAl:1992} to generate a Kolmogorov phase screen. It can be used
for any wavefront structure function. It allows us to very efficiently
apply the inverse of the phase covariance matrix to any vector.

We show that this fractal operator is also very efficient when used as a
preconditioner. It allows the wavefront reconstruction to be iteratively
computed in a space of statistically independent modes. We additionally use
a classical Jacobi preconditioner, or a new ``optimal diagonal
preconditioner'' to further improve the convergence.  The number of
iterations is $\lesssim10$ for a full wavefront reconstruction
whatever the size of the system, with a number of floating point operations
$\sim 34\times{}\Nparam$ per iteration.  The method is therefore globally
$\ScaleAs(\Nparam)$.

In the following, we first derive the analytical expression for the minimum
variance restored wavefront and the equations to be solved. We then
introduce the fractal operator allowing fast computation of the
regularization term in an iterative method such as conjugate gradients.
We then propose two fast preconditioners to further speed up the iterative
algorithm. We finally use numerical simulations to test the performances of \FRIM.

\section{Minimum variance solution}

\subsection{Model of data}

We assume that the wavefront sensor provides measurements of spatial
derivatives (slopes or curvatures) of the phase, which are linearly related
to the wavefront seen by the sensor:
\begin{equation}
  \V{d}=\M{S}\cdot\V{w}+\V{n}
  \label{eq:data-model}
\end{equation}
where $\V{d}\in\mathbb{R}^\Ndata$ is the \emph{data} vector provided by the
sensor, $\V{w}\in\mathbb{R}^\Nparam$ is the vector of sampled wavefront
values, $\M{S}\in\mathbb{R}^{\Ndata\times\Nparam}$ is the sensor response
matrix and $\V{n}\in\mathbb{R}^\Ndata$ accounts for the noise and model
errors.  This equation is general as long as the wavefront sensor is
linear.  As a typical case, we will however consider a Shack-Hartmann
wavefront sensor with Fried geometry \cite{Fried_1977} in our simulations
and for the evaluation of the efficiency of the algorithms.

\subsection{Optimal wavefront reconstructor}

The estimation of the wavefront $\V{w}$ given the data $\V{d}$ is an
inverse problem which must be solved using proper regularization in
order to improve the quality of the solution while avoiding noise
amplification or ambiguities due to missing data
\cite{Thiebaut:2005:Cargese}. In order to keep the problem as simple as
possible, we first introduce the requirement that the solution be a linear
function of the data, \ie the restored wavefront satisfies:
\begin{equation}
  \tilde{\V{w}}\bydef\M{R}\cdot\V{d}
  \label{eq:linear-solution}
\end{equation}
where $\M{R}$ is the restoration matrix and $\V{d}$ the wavefront sensor
measurements. Some quality criterion is needed to derive the expression for
the restoration matrix $\M{R}$. For instance, we can require that, on
average, the difference between the restored wavefront $\tilde{\V{w}}$ and
the true wavefront $\V{w}$ be as small as possible by minimizing
$\avg{\norm{\tilde{\V{w}}-\V{w}}^{2}}$ where $\avg{\cdot}$ denotes the
expected value of its argument. It is interesting to note that minimizing
(on average) the variance of the residual wavefront yields the optimal
Strehl ratio \cite{Herrmann_1992} since:
\begin{equation}
  \mathrm{SR}\simeq\exp\left(-\frac{1}{\mathcal{A}}\,
  \int_{\mathrm{pupil}}\left[\tilde{w}\left(\V{r}\right)-w\left(\V{r}\right)
    \right]^{2}\,\D{\V{r}}\right)
  \label{eq:strehl-ratio}
\end{equation}
where $\V{r}$ is the position in the pupil, $\mathcal{A}$ is the area of
the pupil and $w\left(\V{r}\right)$ is the wavefront phase in radian units.
The \emph{best} reconstruction matrix according to our criterion then
satisfies:
\begin{equation}
  \M{R}^{\dagger}=\argmin_{\M{R}}
  \avg{\norm{\M{R}\cdot\V{d}-\V{w}}^{2}} \,.
  \label{eq:R-best-def}
\end{equation}
Accounting for the facts that the wavefront $\V{w}$ and the errors $\V{n}$
are uncorrelated and have zero means, \ie $\avg{\V{n}}=0$ and
$\avg{\V{w}}=0$, the minimum variance reconstructor expands as
\cite{Tarantola-2005-inverse_problem_theory}:
\begin{equation}
  \label{eq:best-R-1}
  \M{R}^{\dagger} = \Cprior\cdot\M{S}\T\cdot
  \left(\M{S}\cdot\Cprior\cdot\M{S}\T + \Cerror\right)^{-1}\,,
\end{equation}
where $\Cerror\bydef\avg{\V{n}\cdot\V{n}\T}$ is the covariance matrix of
the errors and $\Cprior\bydef\avg{\V{w}\cdot\V{w}\T}$ is the \emph{a
priori} covariance matrix of the wavefront.  Applying this reconstructor to
the data $\V{d}$ requires solving a linear problem with as many equations
as there are measurements.  Generally, wavefront sensors provide more
measurements than wavefront samples (about twice as many for a
Shack-Hartmann or a curvature sensor).  Fortunately, from the following
obvious identities \cite{TarantolaValette:1982}:
\begin{align}
  \M{S}\T\cdot\Cerror^{-1}\cdot\M{S}\cdot\Cprior\cdot\M{S}\T + \M{S}\T
  &= \M{S}\T\cdot\Cerror^{-1}\cdot\left(
       \M{S}\cdot\Cprior\cdot\M{S}\T + \Cerror
     \right)
  \notag \\
  &= \left(
       \M{S}\T\cdot\Cerror^{-1}\cdot\M{S} + \Cprior^{-1}
     \right)\cdot\Cprior\cdot\M{S}\T
  \,,\notag
\end{align}
we can rewrite the optimal reconstructor in Eq.~(\ref{eq:best-R-1}) as:
\begin{equation}
  \label{eq:best-R-2}
  \M{R}^{\dagger} =
  \left(\M{S}\T\cdot\Cerror^{-1}\cdot\M{S} + \Cprior^{-1}\right)^{-1}
  \cdot\M{S}\T\cdot\Cerror^{-1}
\end{equation}
which involves solving just as many linear equations as there are wavefront
samples. The linear reconstructor defined in Eq.~(\ref{eq:best-R-2}) is the
expression to be preferred in our case.

\subsection{Links with other approaches}

Using Eq.~(\ref{eq:best-R-2}) for the reconstructor,
the minimum variance restored wavefront is given by:
\begin{displaymath}
  \V{w}^{\dagger}
  \bydef \M{R}^{\dagger}\cdot\V{d}
  = \left(\M{S}\T\cdot\Cerror^{-1}\cdot\M{S}+\Cprior^{-1}\right)^{-1}
  \cdot\M{S}\T\cdot\Cerror^{-1}\cdot\V{d}
\end{displaymath}
which is also the solution of the quadratic problem:
\begin{displaymath}
  \V{w}^{\dagger} = \argmin_{\V{w}}\left\{
  \left(\M{S}\cdot\V{w} - \V{d}\right)\T\cdot\Cerror^{-1}\cdot
  \left(\M{S}\cdot\V{w} - \V{d}\right)
  + \V{w}\T\cdot\Cprior^{-1}\cdot\V{w}\right\}
\end{displaymath}
where
$\left(\M{S}\cdot\V{w}-\V{d}\right)\T\cdot\Cerror^{-1}\cdot\left(\M{S}\cdot\V{w}-\V{d}\right)$
is the so-called $\chi^{2}$ which measures the discrepancy between the data
and their model and $\V{w}\T\cdot\Cprior^{-1}\cdot\V{w}$ is a Tikhonov
regularization term which enforces \emph{a priori} covariance of the
unknowns.  Thus \Eq{eq:best-R-2} is also the result of the maximum a
posteriori (MAP) problem.  Here, the usual hyper-parameter is hidden in
$\Cprior$ which is proportional to $(D/r_0)^{5/3}$, where $r_0$ is the
Fried parameter \cite{Fried_1965}.  As already noted by
other authors (see \eg Rousset \cite{Rousset_1993}), the minimum variance
estimator is directly related to Wiener optimal filtering.

Actual adaptive optics systems make use of some expansion of the wavefront
on a basis of modes, regularization being achieved by setting the
ill-conditioned modes to zero. This technique is similar to truncated
singular value decomposition (TSVD) \cite{Rousset_1993}. Since truncation
results in aliasing, we expect that the MAP solution will be a better
approximation to the wavefront.


\subsection{Iterative Method}

\begin{figure}
  \centerline{
    \fbox{
      \begin{minipage}{70mm}
	\begin{flushleft}
	  initialisation: \\
	  \quad compute $\V{r}_{0}=\V{b}-\M{A}\cdot\V{x}_{0}$
	  for some initial guess $\V{x}_{0}$ \\
	  \quad let $k=0$ \\
	  until convergence do \\
	  \quad solve $\M{M}\cdot\V{z}_{k}=\V{r}_{k}$ for $\V{z}_{k}$
	        \hfill(apply preconditioner) \\
	  \quad $\rho_{k}=\V{r}_{k}\T\cdot\V{z}_{k}$ \\
	  \quad if $k=0$, then \\
	  \quad \quad $\V{p}_{k}=\V{z}_{k}$ \\
	  \quad else \\
	  \quad \quad $\V{p}_{k}=\V{z}_{k}+(\rho_{k}/\rho_{k-1})\,\V{p}_{k-1}$ \\
	  \quad endif \\
	  \quad $\V{q}_{k}=\M{A}\cdot\V{p}_{k}$ \\
	  \quad $\alpha_{k}=\rho_{k}/(\V{p}_{k}\T{\cdot}\V{q}_{k})$
	  \hfill(optimal step size) \\
	  \quad $\V{x}_{k+1}=\V{x}_{k}+\alpha_{k}\,\V{p}_{k}$ \\
	  \quad $\V{r}_{k+1}=\V{r}_{k}-\alpha_{k}\,\V{q}_{k}$ \\
	  \quad $k\leftarrow k+1$ \\
	  done
	\end{flushleft}
      \end{minipage}
    }
  }
  \caption{\label{fig:conjugate-gradient-algorithm}Preconditioned conjugate
    gradient algorithm for solving $\M{A}{\cdot}\V{x}=\V{b}$ where $\M{A}$
    is a symmetric positive definite matrix and $\M{M}$ is a
    preconditioner.  The unpreconditioned version of the algorithm is
    simply obtained by taking $\M{M}=\M{I}$, hence $\V{z}_{k}=\V{r}_{k}$.}
\end{figure}

The optimal wavefront can be computed in different ways. For
instance, the matrix $\M{R}$ can be computed once, using
\Eq{eq:best-R-1} or \Eq{eq:best-R-2}, and then applied to every data set
$\V{d}$.  Since it requires the numerical inversion of an
$\Nparam\times{}\Nparam$ matrix, the direct computation of $\M{R}$ scales
as $\ScaleAs(\Nparam^{3})$ operations \cite{NumericalRecipes}. The
reconstructor $\M{R}$ is a $\Nparam\times\Ndata$ matrix and is not sparse
in practice. Hence, the storage of $\M{R}$ requires $\Ndata\,\Nparam
\approx 2\,\Nparam ^2$ floating point numbers and computing
$\M{R}\cdot\V{d}$ requires $\approx2\,\Ndata\,\Nparam \approx
4\,\Nparam^{2}$ floating point operations.  For large numbers of degrees of
freedom $\Nparam\propto (D/r_0)^2$, the computer time spent by the
matrix-vector multiplication can be too long for real time applications.
Moreover the memory requirement (\eg for $\Nparam\simeq 10^4$,
$1.5\,\mathrm{Gb}$ of memory are needed to store $\M{R}$) may be such that
memory page faults dominate the computation time of matrix-vector
multiplication.

In order to avoid the direct matrix inversion and the matrix-vector
product required by the explicit computation of $\M{R}$, we use an
iterative method to solve the linear system
\begin{equation}
  \label{eq:linear-system}
  \left(\M{S}\T\cdot\Cerror^{-1}\cdot\M{S}+\Cprior^{-1}\right)\cdot\V{w}
  = \M{S}\T\cdot\Cerror^{-1}\cdot\V{d}
\end{equation}
which leads to the optimal wavefront $\V{w}$ for every data set $\V{d}$.
For the purpose of the discussion, \Eq{eq:linear-system} can be put in a
more generic form:
\begin{equation}
  \label{eq:Ax=b}
  \M{A}\cdot\V{x}=\V{b}
\end{equation}
where, in the case of \Eq{eq:linear-system}, $\V{x}=\V{w}$ and:
\begin{equation}
  \label{eq:lhs-matrix}
  \M{A}=\M{S}\T\cdot\Cerror^{-1}\cdot\M{S} + \Cprior^{-1}
\end{equation}
is the so-called left hand side matrix, whereas
\begin{equation}
  \label{eq:rhs-vector}
  \V{b}=\M{S}\T\cdot\Cerror^{-1}\cdot\V{d}
\end{equation}
is the so-called right hand side vector.

Barrett \etal \cite{templates} have reviewed a number of iterative
algorithms for solving linear systems like (\ref{eq:Ax=b}).  An advantage
of these methods is that they do not explicitly require the matrix $\M{A}$;
it is sufficient to be able to compute the product of matrix $\M{A}$ (or
its transpose) with any given vector.  The iterative algorithm therefore
fully benefits from the possibility to compute the matrix-vector products
in much less than $\ScaleAs(\Nparam^2)$ operations when $\M{A}$ is sparse
or has some special structure.  This is particularly relevant in our case
since applying $\M{A}$ can be achieved by matrix-vector products very fast to compute
as shown in Sect.~\ref{sec:sensor-model} and
Sect.~\ref{sec:regularization-term}.  The drawback of iterative methods is
that the computational burden scales as the number of iterations required
to approximate the solution with sufficient precision.  In the worst
case, the number of iterations can theoretically be as high as the number
of unknowns $\Nparam$ \cite{Nocedal_Wright-2006-numerical_optimization,
  NumericalRecipes}.  In practice and because of numerical rounding errors,
ill-conditioning of the system in \Eq{eq:linear-system} can result in a
much higher number of iterations, even on small systems.  This problem can
however be greatly reduced by means of a good preconditioner
\cite{templates,Nocedal_Wright-2006-numerical_optimization}.

By construction, $\M{A}$ given by \Eq{eq:lhs-matrix} is a symmetric
positive definite matrix and the conjugate gradient (CG) \cite{templates}
is the iterative method of choice to solve the system in
Eq.~(\ref{eq:Ax=b}). Figure~\ref{fig:conjugate-gradient-algorithm} shows
the steps of the CG algorithm to solve the system $\M{A}\cdot\V{x}=\V{b}$.
This method is known to have a super-linear rate of convergence
\cite{Nocedal_Wright-2006-numerical_optimization}, and can be accelerated
by using a proper preconditioner $\M{M}\approx\M{A}$ for which solving
$\M{M}\cdot\V{z}=\V{r}$ for $\V{z}$ (with $\V{r}=\V{b}-\M{A}\cdot\V{x}$) is much cheaper than solving
\Eq{eq:Ax=b} for $\V{x}$. The preconditioner can also be directly specified
by its inverse $\M{Q}=\M{M}^{-1}$ such that $\M{Q}\approx\M{A}^{-1}$ and
then $\V{z}=\M{Q}\cdot\V{r}$ in the CG algorithm. Without a preconditioner,
taking $\M{M}=\M{Q}=\M{I}$, where $\M{I}$ is the identity matrix, yields
the unpreconditioned version of the CG algorithm. In
Sect.~\ref{sec:preconditioning} we investigate various means to obtain an
effective preconditioner for the wavefront reconstruction problem.

In the remainder of this section, we derive means to quickly compute the
dot product with the matrix $\M{A}$ in \Eq{eq:lhs-matrix}.  To that end, we
consider separately the Hessian matrix $\M{S}\T\cdot\Cerror^{-1}\cdot\M{S}$
of the likelihood term and that of the regularization term $\Cprior^{-1}$.

\subsection{Computation of the likelihood term}
\label{sec:sensor-model}

\begin{figure}
  \centering
  \includegraphics[height=40mm]{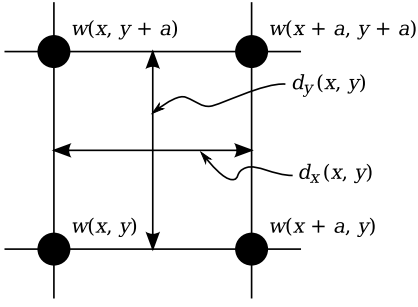}
  \caption{\label{fig:wfs}Wavefront sensor with Fried geometry as used
    for our simulations. The black circles stand for phase samples
    $w(x,y)$, at the corners of the square subapertures of size $a$. This
    model is exact if we assume that the wavefront at any point in the
    pupil is obtained from a bilinear interpolation of phase samples at the
    corner of the subapertures.}
\end{figure}

Most adaptive optics systems use either a Shack-Hartmann sensor which
provides measurements of the local gradient of the wavefront or a curvature
sensor which measures the local curvature of the wavefront
\cite{Roddier_1999}. Since such sensors probe local spatial derivatives of
the wavefront, their response can be approximated by local finite
differences which yields a very sparse linear operator $\M{S}$.  Though
some non-sparse matrix terms can appear due to tilt indetermination with
laser guide stars or to take account of natural guide star tip/tilt
sensors. Owing to the low rank of these modes, sparse matrix models can
still be applied \cite{Ellerbroek_2002}.  Thus, denoting $\Ndif$ the number
of wavefront samples required to compute the local finite differences, only
$\approx \Ndata \times \Ndif$ out of $\Ndata\times\Nparam$ coefficients of
$\M{S}$ are non-zero.  For instance, \Fig{fig:wfs} shows the Fried
geometry of the Shack-Hartmann sensor model \cite{Fried_1977} which we used
in our numerical simulations. The error free slopes are related to the
wavefront by:
\begin{equation}
  \label{eq:Fried-model}
  \begin{split}
  d_\RM{x}(x,y) & = \OneOverTwo\,\bigl[
    w(x + a, y + a) + w(x + a, y)\\
    &\quad -\ w(x, y + a) - w(x, y)
    \bigr]\\
  d_\RM{y}(x,y) & = \OneOverTwo\,\bigl[
    w(x + a, y + a) - w(x + a, y)\\
    &\quad +\ w(x, y + a) - w(x, y)
    \bigr]  
  \end{split}
\end{equation}
where $(x,y)$ are the pupil coordinates, $d_\RM{x}$ and $d_\RM{y}$ are the
slopes along the $x$ and $y$ directions and $a$ is the sampling step. Hence
$\Ndif=4$, in our case, whatever the number of degrees of freedom.
Besides, to a good approximation, wavefront sensors provide uncorrelated
measurements \cite{Roddier_1999}, hence the covariance matrix $\Cerror$ of
the errors can be taken as a diagonal matrix:
\begin{equation}
  \Cerror \approx \diag\bigl(\Var(\V{n})\bigr)
\end{equation}
where $\Var(\V{n})$ is the vector of noise and error variances. Since
$\Cerror$ is diagonal, its inverse $\Cerror^{-1}$ is diagonal and trivial
to compute.  Finally, the matrices $\M{S}$ and $\Cerror^{-1}$ are sparse
and the dot product by $\M{S}\T\cdot\Cerror^{-1}\cdot\M{S}$ can
be therefore computed in $\ScaleAs(\Nparam)$ operations.

\subsection{Fast estimation of the regularization term}
\label{sec:regularization-term}

Unlike $\Cerror$ and $\Cerror^{-1}$, neither $\Cprior$ nor $\Cprior^{-1}$
is sparse. We introduce here a way to derive an approximation for
$\Cprior^{-1}$ so that it can be applied to a vector with a small number of
operations.

We first consider the following decomposition of $\Cprior$:
\begin{equation}
  \label{eq:Cprior-factorization}
  \Cprior=\M{K}\cdot\M{K}\T \,
\end{equation}
where $\M{K}$ is a square invertible matrix. Since $\Cprior$ is positive
definite, there exists a number of possibilities for such a factorization:
Cholesky decomposition \cite{NumericalRecipes, Ellerbroek_2002,
  Gilles_2003b}, spectral factorization, \etc We then use this
decomposition to define new variables $\V{u}$ based on the wavefront
$\V{w}$:
\begin{equation}
  \label{eq:phase-gen-def}
  \V{u}\bydef\M{K}^{-1}\cdot\V{w} \, .
\end{equation}
The expected value of $\V{u}$ is:
$\avg{\V{u}}=\M{K}^{-1}\cdot\avg{\V{w}}=\V{0}$ and its covariance matrix
therefore satisfies:
\begin{align}
  \M{C}_{\V{u}}
  &= \avg{\V{u}\cdot\V{u}\T}
   = \M{K}^{-1}\cdot\avg{\V{w}\cdot\V{w}\T}\cdot\M{K}\mT
   \nonumber\\
  &= \M{K}^{-1}\cdot\Cprior\cdot\M{K}\mT
   = \M{I}
   \nonumber\,,
\end{align}
which shows that the new variables are independent and identically
distributed following a normal law: $\V{u}\sim\mathcal{N}(\V{0},\M{I})$.
This gives rise to a method for generating wavefronts since from a set
$\V{u}$ of $\Nparam$ independent random values following a normal law,
taking $\V{w}=\M{K}\cdot\V{u}$ yields a random wavefront with the expected
covariance. Finally, using this re-parametrization, it is possible to
rewrite the regularization term as:
\begin{equation}
  \label{eq:regul_with_u}
  \V{w}\T\cdot\Cprior^{-1}\cdot\V{w}
  = \V{w}\T\cdot\M{K}\mT\cdot\M{K}^{-1}\cdot\V{w}
  = \Norm{\M{K}^{-1}\cdot\V{w}}_{2}^{2}
  = \norm{\V{u}}_{2}^{2}\,.
\end{equation}
Then, depending on whether the problem is solved for the wavefront samples
$\V{w}$ or for the so-called \emph{wavefront generators} $\V{u}$ (\cf
equations (\ref{eq:CG_for_w}) and (\ref{eq:CG_for_u}) in
Sect.~\ref{sec:results}), each conjugate gradient iteration would be cheap
to compute providing either (i) operators $\M{K}^{-1}$ and $\M{K}\mT$ are
fast to apply, or (ii) operators $\M{K}$ and $\M{K}\T$ are fast to apply.

A comparable re-parametrization has been introduced by Roddier
\cite{Roddier-1990-wavefront_simulation} for generating turbulent
wavefronts using a Zernike expansion of randomly weighted
Karhunen-Lo\`eve functions. This however requires to diagonalize a huge
$\Nparam\times{}\Nparam$ matrix $\Cprior$, a procedure that costs at least
$\ScaleAs(\Nparam^3)$ operations, and would give a \emph{slow} operator
$\M{K}$ (or $\M{K}^{-1}$) taking $\ScaleAs(\Nparam^2)$ operations to apply.

Exploiting the fractal structure of turbulent wavefronts, Lane \etal
\cite{LaneEtAl:1992} have derived a fast method to generate wavefronts by a
mid-point algorithm. In what follows, we show that their method amounts to
approximating the effect of operator $\M{K}$ in $\ScaleAs(\Nparam)$
operations and we derive algorithms to apply the corresponding
$\M{K}^{-1}$, $\M{K}\T$, and $\M{K}\mT$ operators that also take
$\ScaleAs(\Nparam)$ operations. We propose to use these so-called
\emph{fractal operators} for fast computation of the regularization and
also as effective pre-conditioners to speed-up the conjugate-gradient
iterations.

\section{Fractal operators}
\label{sec:fractal_operators}

\subsection{Principle and structure function}

The mid-point algorithm \cite{LaneEtAl:1992} starts at the largest scales
of the wavefront and step-by-step builds smaller scales by interpolating
the wavefront values at the previous scale and by adding a random value
with a standard deviation computed so that the new wavefront values and
their neighbors have the expected structure function.  Using $\M{K}_j$ to
denote the linear operator which generates the wavefront values at the $j$-th
scale, the linear operator $\M{K}$ can be factorized as:
\begin{equation}
  \label{eq:K_factorized}
  \M{K} = \M{K}_1\cdot\M{K}_2\cdot\ldots\cdot\M{K}_p
\end{equation}
where $p$ is the number of scales, $\M{K}_p$ generates the 4 outermost
wavefront values and $\M{K}_1$ generates the wavefront values at the finest
scale.  The original mid-point algorithm cannot be used directly for our
needs because it is not invertible.  In this section, we reconsider the
mid-point algorithm to derive new expressions for the $\M{K}_j$'s such that
they are sparse, invertible and such that their inverses are also sparse.

The structure function of the wavefront is the expected value of the
quadratic difference between two phases of a turbulent wavefront:
\begin{equation}
  \bigl\langle\bigl[w(\V{r}_{i}) - w(\V{r}_{j})\bigr]^2\bigr\rangle
  = f\bigl(\abs{\V{r}_{i} - \V{r}_{j}}\bigr) \, ,
\end{equation}
where, \eg:
\begin{equation}
  \label{eq:f-def}
  f(r) = 6.88\times\bigl(r / r_0\bigr)^{5/3} \, ,
\end{equation}
for a turbulent wavefront obeying Kolmogorov's law.  The structure function
is stationary (shift-invariant) and isotropic since it only depends on the
distance $\abs{\V{r}_{i}-\V{r}_{j}}$ between the considered positions
$\V{r}_{i}$ and $\V{r}_{j}$ in the wavefront.  From the structure function,
we can deduce the covariance of the wavefront between two positions in the
pupil:
\begin{equation}
  \label{eq:covariance}
  C_{i,j}
  = \avg{w_{i}\, w_{j}}
  = \OneOverTwo\,\bigl(\sigma_{i}^{2} + \sigma_{j}^{2}
     - f_{i,j}\bigr)
\end{equation}
with $w_{i}=w(\V{r}_i)$ the wavefront phase at position $\V{r}_i$,
$\sigma_{i}^{2}=\Var(w_{i})$, and $f_{i,j}=f(\abs{\V{r}_{i} - \V{r}_{j}})$
the structure function between wavefront samples $i$ and $j$.  The
wavefront variances (thus the covariance) are not defined for pure
Kolmogorov statistics but can be defined by other models of the
turbulence such as the von K\'arm\'an model.  Nevertheless, any structure
function $f$ can be used by our algorithm: in case the variance is
undefined, we will show that the $\sigma_{i}^{2}$'s appear as free
parameters and that choosing suitable variance values is not a problem.

\subsection{Generation of outermost values}
\label{sec:outermost-scale}

The first point to address is the initialization of the mid-point
recursion, that is the generation of the four outermost corner values.
Lane \etal \cite{LaneEtAl:1992} used 6 random values to generate the 4
initial corners.  It is however required to use exactly the same number of
random values $\V{u}$ as there are wavefront samples in $\V{w}$ otherwise
the corresponding linear operator $\M{K}$ cannot be invertible.  This is
possible by slightly modifying their original algorithm.

\begin{figure}
  \centerline{\includegraphics[height=30mm]{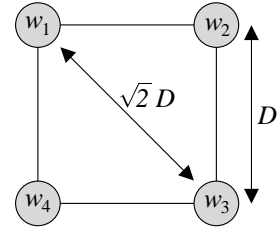}}
  \caption{\label{fig:fractal-step0}The four initial values for wavefront
  generation, at the corners of the support.}
\end{figure}

The four initial wavefront values (\Fig{fig:fractal-step0}) have the
following covariance matrix:
\begin{displaymath}
  \M{C}_\RM{out} = \Matrix{cccc}{
    c_{0} & c_{1} & c_{2} & c_{1}\\
    c_{1} & c_{0} & c_{1} & c_{2}\\
    c_{2} & c_{1} & c_{0} & c_{1}\\
    c_{1} & c_{2} & c_{1} & c_{0}\\
  }
  \quad\text{with}\quad
  \left\{\begin{array}{rcl}
    c_{0} &=& \sigma^{2} \\
    c_{1} &=& \sigma^{2} - f(D)/2 \\
    c_{2} &=& \sigma^{2} - f(\SqrtTwo\,D)/2 \\
  \end{array}\right.
\end{displaymath}
where $\sigma^{2}$ is the variance (assumed to be the same) of the four
initial phases and where $D$ is the distance between points 1 and 2 (see
Fig.~\ref{fig:fractal-step0}). Having the same variances $\sigma^2$ for the
four outermost wavefront phases seems natural since none of these points
plays a particular role. For the four outer wavefront samples, the matrix of
eigenvectors of $\M{C}_\RM{out}$ is:
\renewcommand{\TmpI}[1]{{\scriptstyle #1}}
\renewcommand{\TmpII}[4]{\TmpI{#1}&\TmpI{#2}&\TmpI{#3}&\TmpI{#4}\\}
\renewcommand{\TmpIII}{\phantom{-}}
\renewcommand{\TmpIV}{1/\sqrt{2}}
\begin{displaymath}
  \M{Z}_\RM{out} = \Matrix{crcc}{
    \TmpII{ 1/2 }{ -1/2 }{   0           }{ \TmpIII\TmpIV}
    \TmpII{ 1/2 }{  1/2 }{    -\TmpIV    }{ 0 }
    \TmpII{ 1/2 }{ -1/2 }{   0           }{ -\TmpIV}
    \TmpII{ 1/2 }{  1/2 }{ \TmpIII\TmpIV }{ 0 }
  }
\end{displaymath}
Note that the eigenvectors (columns) defined on these four samples are (in
order) \emph{piston}, \emph{waffle} \cite{MacMartin_2003}, \emph{tip} and
\emph{tilt}. The eigenvalues are:
\begin{displaymath}
  \BS{\lambda}_\RM{out}
  = \Matrix{c}{
    c_{0} + 2\,c_{1} + c_{2} \\
    c_{0} - 2\,c_{1} + c_{2} \\
    c_{0} - c_{2} \\
    c_{0} - c_{2} \\
  }
  = \Matrix{c}{
    4\,\sigma^2 - f(D) - f(\SqrtTwo\,D)/2 \\
    f(D) - f(\SqrtTwo\,D)/2 \\
    f(\SqrtTwo\,D)/2 \\
    f(\SqrtTwo\,D)/2 \\
  }\,.
\end{displaymath}

In the case of pure Kolmogorov statistics, $\sigma^2$ must be chosen so that
$\M{K}$ is invertible.  This is achieved if the eigenvalue of the
\emph{piston}-like mode is strictly positive, hence:
\begin{displaymath}
  \sigma^2 > f(D)/4 + f(\SqrtTwo\,D)/8 \, .
\end{displaymath}
We have chosen $\sigma^{2}$ so that the smallest covariance, which is
$c(\SqrtTwo\,D)$ between the most remote points, is exactly zero:
\begin{equation}
  \label{eq:sigma_with_Kolmogorov}
  \sigma^2=\OneOverTwo\,f(\SqrtTwo\,D)\ .
\end{equation}
Of course, when a von K\'arm\'an model of turbulence is chosen, both
$\sigma^2$ and $f$ are fixed by the model; \Eq{eq:sigma_with_Kolmogorov} is
to be used only for the Kolmogorov case.

A possible expression for the operator $\M{K}_\RM{out}$, such that
$\M{C}_\RM{out}=\M{K}_\RM{out}\cdot\M{K}_\RM{out}\T$, is:
\begin{align}
  \M{K}_\RM{out} &= \frac{1}{2} \, \Matrix{rrrr}{
    a & -b & -c &  0\\
    a &  b &  0 & -c\\
    a & -b &  c &  0\\
    a &  b &  0 &  c
  } \, ,  \label{eq:k_out} \\
  \text{with:\ }
    a &= {\textstyle\sqrt{4\,\sigma^2 - f(D) - f(\SqrtTwo\,D)/2}}\,, \notag\\
    b &= {\textstyle\sqrt{f(D) - f(\SqrtTwo\,D)/2}}\,, \notag\\
    c &= {\textstyle\sqrt{f(\SqrtTwo\,D)}}\,, \notag
\intertext{from which $\M{K}_\RM{out}^{-1}$ is:}
  \M{K}_\RM{out}^{-1} &= \frac{1}{2} \, \Matrix{rrrr}{
     1/a &  1/a &  1/a & 1/a\\
    -1/b &  1/b & -1/b & 1/b\\
    -2/c &    0 &  2/c &   0\\
       0 & -2/c &    0 & 2/c
  } \, .  \label{eq:k_out_inv}
\end{align}
The operator $\M{K}_p$ in Eq.~(\ref{eq:K_factorized}) is obtained simply
from $\M{K}_\RM{out}$.  Indeed, $\M{K}_p$ is essentially the identity
matrix except for 16 non-zero coefficients corresponding to the outermost
corners and which are given by $\M{K}_\RM{out}$.  The same rules yield
$\M{K}_p^{-1}$ from $\M{K}_\RM{out}^{-1}$.

\subsection{Generation of wavefront samples at smaller scales}
\label{sec:generation_smaller_scales}

\begin{figure}
  \centerline{\includegraphics[height=70mm]{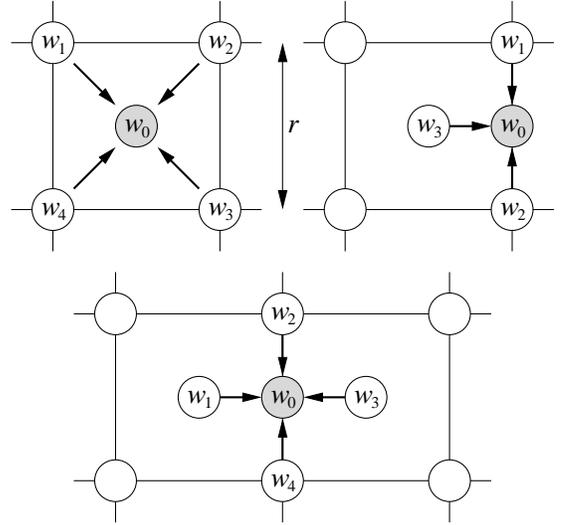}}
  \caption{\label{fig:fractal-step1}Wavefront refinement. To generate a
    grid with cell size $r/2$, new values (in gray) are generated from
    wavefront values (in white) of a grid with cell size equal to $r$. Top
    left: new value from 4 values $r/\sqrt{2}$ apart. Top right: new edge
    value from 3 values $r/2$ apart. Bottom: new value from 4 values $r/2$
    apart.}
\end{figure}

Given the wavefront with a sampling step $r$, the mid-point algorithm
generates a refined wavefront with a sampling of $r/2$ using a perturbed
interpolation:
\begin{equation}
  \label{eq:mid-point}
  w_{0} = \alpha_{0}\,u_{0} + \sum_{j=1}^{\Nint}\alpha_{j}\,w_{j}
\end{equation}
where $w_{0}$ is the wavefront value at the mid-point position,
$u_{0}\sim\NormalLaw\left(0,1\right)$ is a normally distributed random
value and $\Nint$ is the number of wavefront samples from the previous
scale which are used to generate the new sample (see
\Fig{fig:fractal-step1}). Equation (\ref{eq:mid-point}) comes from a
generalization of the principle of the original mid-point algorithm. Since
we proceed from the largest scale to smaller ones, all the operations can
be done \emph{in-place}: the value of $w_{0}$ computed according to
\Eq{eq:mid-point} replacing that of $u_{0}$. In other words, the input and
output vectors, $\V{u}$ and $\V{w}$, can share the same area of the
computer memory. It is then immediately apparent that a random wavefront
computed by this algorithm scales as
$\ScaleAs(\Nint\times{}\Nparam)=\ScaleAs(\Nparam)$ since the number of
neighbors $\Nint\sim4$ does not depend on the number of wavefront samples
$\Nparam$.

The $\Nint+1$ scalars $\alpha_{j}$ have to be adjusted so that the
structure function between $w_{0}$ and any of the $w_{i=1,\ldots,\Nint}$
matches the turbulence statistics:
\begin{align}
  f_{i,0}
  &\bydef \avg{(w_{0}-w_{i})^{2}} \notag\\
  &= \alpha_{0}^{2} + \sum_{j=1}^{\Nint}\alpha_{j}\,f_{i,j}
      - \DoubleSumNint{j}{k}
        \alpha_{j}\,\alpha_{k}\,f_{j,k} \nonumber\\
  &\quad + \left(1 - \sum_{k=1}^{\Nint}\alpha_{k}\right)\,
        \left(\sigma_{i}^{2} - \sum_{j=1}^{\Nint}\alpha_{j}
        \,\sigma_{j}^{2}\right) \,.
	\label{eq:struct-fn-expand}
\end{align}
Note that, to obtain this equation, we have accounted for the fact that
since $u_{0}\sim\mathcal{\Nparam}\left(0,1\right)$ and
$w_{j=1,\ldots,\Nint}$ are uncorrelated, then $\avg{u_{0}^{2}}=1$ and
$\avg{u_{0}\,w_{j=1,\ldots,\Nint}}=0$.  The
system~(\ref{eq:struct-fn-expand}) gives $\Nint$ equations, whereas there
are $\Nint+1$ unknown parameters $\{\alpha_{0},\ldots,\alpha_{\Nint}\}$: an
additional constraint is needed.

In the original mid-point algorithm, Lane \etal \cite{LaneEtAl:1992} choose to
normalize the sum of the interpolation coefficients and use the constraint
that $\sum_{j=1}^{\Nint}\alpha_{j}=1$.  In that case,
\Eq{eq:struct-fn-expand} simplifies and the coefficients are obtained by
solving:
\begin{equation}
  \begin{split}
    f_{i,0} = \alpha_{0}^{2} + \sum_{j=1}^{\Nint}\alpha_{j}\,f_{i,j}
    - \DoubleSumNint{j}{k}\alpha_{j}\,\alpha_{k}\,f_{j,k} \\
    \text{s.t.}\quad\sum_{j=1}^{\Nint}\alpha_{j}=1\,.
  \end{split}
  \label{eq:orig-alpha-system}
\end{equation}
Note that all the variances $\sigma_{j}^{2}$ are implicit with this
constraint.

We consider here another constraint which is to have the same variance,
say $\sigma^{2}$, for all the wavefront samples. In other words, we
consider a wavefront with stationary (shift-invariant) statistical
properties. This is justified by our objective to reconstruct phase
corrugations in several layers for atmospheric tomography. Indeed, since
the beams coming from different directions in the field of view are not
superimposed in the layers, this condition allows the wavefront statistics
to remain the same for all the beams. With this choice, the additional
equation is provided by $\avg{w_{0}^{2}}=\sigma^{2}$ and the interpolation
coefficients $\{\alpha_{0},\ldots,\alpha_{\Nint}\}$ are obtained by solving
the system of $\Nint+1$ equations:
\begin{displaymath}
  \begin{split}
    f_{i,0} &= \alpha_{0}^{2} + \sum_{j=1}^{\Nint}\alpha_{j}\,f_{i,j}
  - \DoubleSumNint{j}{k}
    \alpha_{j}\,\alpha_{k}\,f_{j,k} \\
  &\quad + \sigma^{2}\,\left(1-\sum_{j=1}^{\Nint}\alpha_{j}\right)^{2}
  \quad\text{for~} i=1,\ldots,\Nint\\
  \sigma^{2} &= \alpha_{0}^{2}
  + \sigma^{2}\,\left(\sum_{j=1}^{\Nint}\alpha_{j}\right)^{2}
  -\DoubleSumNint{j}{k}
  \alpha_{j}\,\alpha_{k}\, f_{j,k}\,.
  \end{split}
\end{displaymath}
The system can be further simplified to:
\begin{equation}
  \begin{split}
    &\!\!\sum_{j=1}^{\Nint}\left(2\,\sigma^2 - f_{i,j}\right)\,\alpha_{j}
    = 2\,\sigma^2 - f_{i,0} \quad\text{for~} i=1,\ldots,\Nint\\
    &\alpha_{0}^{2}
    = \left[1 - \left(\sum_{j=1}^{\Nint}\alpha_{j}\right)^{2}\right]\,\sigma^2
    + \DoubleSumNint{j}{k}
    \alpha_{j}\,\alpha_{k}\, f_{j,k}\,,
  \end{split}
  \label{eq:alpha-system}
\end{equation}
where the first $\Nint$ equations form a linear system which must be solved
to obtain the $\alpha_{j=1,\ldots,\Nint}$ and where substituting these
values in the last equation yields the value of $\alpha_0$.  It is worth
noting that by using the covariances instead of the structure function, the
system in \Eq{eq:alpha-system} is equivalent to:
\begin{equation}
  \begin{split}
    &\sum_{j=1}^{\Nint} C_{i,j}\,\alpha_{j} = C_{0,i}
     \quad\text{for~} i=1,\ldots,\Nint\\
    &\alpha_{0}^{2}
    = \sigma^2 - \sum_{j=1}^{\Nint} C_{0,j}\,\alpha_j\,.
  \end{split}
  \label{eq:alpha-system-covar}
\end{equation}

\begin{figure}
  \centering
  \includegraphics[width=80mm,keepaspectratio]{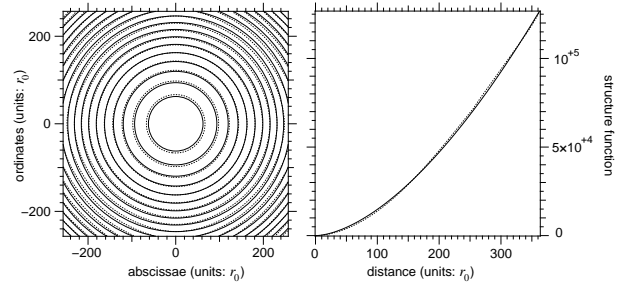}%
  \caption{\label{fig:struct-fn} Structure function.  \emph{Left:} 2D
    isocontours; \emph{right:} 1D profile computed by radial averaging.
    \emph{Solid lines:} Kolmogorov law $6.88\times(r/r_0)^{5/3}$;
    \emph{dotted lines:} average of 1000 structure functions generated with
    the mid-point method.}
\end{figure}

The expressions for the interpolations coefficients for the different cases
illustrated by \Fig{fig:fractal-step1} are derived in
Appendix~\ref{sec:coefs}. To assess the accuracy of the statistics
approximated by the fractal operator, we have computed the structure
function of phase screens $\V{w}$ computed by our implementation of the
mid-point algorithm, \ie as $\V{w}=\M{K}\cdot\V{u}$ with
$\V{u}\sim\NormalLaw\!(\V{0},\M{I})$. Figure~\ref{fig:struct-fn} shows that
the 2D structure function is almost isotropic and demonstrates good
agreement of our approximation to the theoretical law.

%
%

\subsection{The inverse operator}
\label{sec:inverse_operator}


According to the factorization in Eq.~(\ref{eq:K_factorized}), the inverse
of $\M{K}$ is:
\begin{equation}
  \label{eq:inverse-of-K}
  \M{K}^{-1}=\M{K}_p^{-1}\cdot\ldots\cdot\M{K}_2^{-1}\cdot\M{K}_1^{-1}\,.
\end{equation}
In section~\ref{sec:outermost-scale}, the inverse of the outermost operator
$\M{K}_{p}$ has been derived and shown to be sparse --- see
Eq.~(\ref{eq:k_out_inv}).  To compute the $\M{K}_{j}^{-1}$'s for the inner
scales ($j<p$), it is sufficient to solve Eq.~(\ref{eq:mid-point}) for
$u_{0}$, which trivially yields:
\begin{equation}
  \label{eq:invert}
  u_{0} = \frac{1}{\alpha_{0}}\,
  \left(
    w_{0}-\sum_{j=1}^{\Nint}\alpha_{j}\, w_{j}
  \right)\,,
\end{equation}
where $\{ w_{1},\ldots,w_{\Nint}\}$ are the neighbors of $w_{0}$
(Fig.~\ref{fig:fractal-step1}).  Since in Eq.~(\ref{eq:invert}), the
$u_j$'s only depend on the $w_j$'s, the $\M{K}_{j}^{-1}$'s can be applied
in any order.  However, by proceeding from the smallest scales toward the
largest ones as in Eq.~(\ref{eq:inverse-of-K}), the operator $\M{K}^{-1}$
can be performed \emph{in-place}.  This property may be important to avoid
memory page faults and to speed-up the computation.  Finally, from
Eq.~(\ref{eq:k_out}) and Eq.~(\ref{eq:invert}), it is clear that applying
the $\M{K}_{j}^{-1}$'s requires exactly as many operations as for the
$\M{K}_{j}$'s and that computing $\M{K}^{-1}\cdot\V{u}$ requires
$\ScaleAs(\Nparam)$ operations.

\subsection{The transpose operator}

Iterating from the smallest scale to the largest one, it is easy to derive
an algorithm to apply the transpose operator
$\M{K}\T=\M{K}_p\T\cdot\ldots\cdot\M{K}_2\T\cdot\M{K}_1\T$ to a given
vector.  The following algorithm computes $\V{z}=\M{K}\T\cdot\V{v}$ for any
input vector $\V{v}$:
\begin{flushleft}
  \quad copy input vector: $\V{z}\leftarrow\V{v}$ \\
  \quad from the smallest scale to the largest scale, do \\
  \quad \quad for $j=1,\ldots,\Nint$ do \\
  \quad \quad \quad $z_{j}\leftarrow z_{j}+\alpha_{j}\,z_{0}$ \\
  \quad \quad done \\
  \quad \quad $z_{0}\leftarrow\alpha_{0}\,z_{0}$ \\
  \quad done \\
  \quad apply $\M{K}_\RM{out}\T$ at the largest scale of $\V{z}$\\
  \quad return $\V{z}$\\
\end{flushleft}
It is important to note that the loop must be performed \emph{in-place} for
the algorithm to work. From the structure of this algorithm, it is clear
that the multiplication of a vector by the transpose operator is performed
in $\ScaleAs(\Nparam)$ operations.

\subsection{The inverse transpose operator}
\label{sec:inverse_transpose_operator}

The operator
$\M{K}\T=\M{K}_{1}\mT\cdot\M{K}_{2}\mT\cdot\ldots\cdot\M{K}_{p}\mT$
works from the largest scale to the smallest one. The following algorithm
computes $\V{z}=\M{K}\mT\cdot\V{v}$ for any input vector
$\V{v}$:
\begin{flushleft}
  \quad copy input vector: $\V{z}\leftarrow\V{v}$ \\
  \quad apply $\M{K}_\RM{out}\mT$ at the largest scale of $\V{z}$\\
  \quad from the largest scale to the smallest scale, do \\
  \quad \quad $z_{0}\leftarrow z_{0}/\alpha_{0}$ \\
  \quad \quad for $j=1,\ldots,\Nint$ do \\
  \quad \quad \quad $z_{j}\leftarrow z_{j}-\alpha_{j}\,z_{0}$ \\
  \quad \quad done \\
  \quad done \\
  \quad return $\V{z}$\\
\end{flushleft}
Again, the operation can be done \emph{in-place} (the copy of the input
vector $\V{v}$ is only required to preserve its contents if needed), and
the number of operations is $\ScaleAs(\Nparam)$.

\section{Preconditioning}
\label{sec:preconditioning}

Preconditioning is a general means to speed up the convergence of iterative
optimization methods \cite{templates} such as the PCG algorithm described
in \Fig{fig:conjugate-gradient-algorithm}. Preconditioning is generally
introduced as finding an invertible matrix $\M{M}$ such that the spectral
properties of $\M{M}^{-1}\cdot\M{A}$ are more favorable than that of
$\M{A}$ (\ie lower condition number and/or more clustered eigenvalues),
and then the transformed system
\begin{equation}
  \M{M}^{-1}\cdot\M{A}\cdot\V{x} = \M{M}^{-1}\cdot\V{b}
\end{equation}
which has the same solution as the original system $\M{A}\cdot\V{x} =
\V{b}$ can be solved in much fewer iterations.  In this section, we
consider different means for preconditioning the phase restoration problem:
explicit change of variables and diagonal preconditioners.

\subsection{Fractal operator as a preconditioner}
\label{sec:K_as_preconditioner}

Preconditioning is also equivalent to an \emph{implicit} linear change of
variables \cite{Nocedal_Wright-2006-numerical_optimization}: using the
preconditioner $\M{M}=\M{C}\T\cdot\M{C}$ in the algorithm of
Fig.~\ref{fig:conjugate-gradient-algorithm} is the same as using the
(unpreconditioned) conjugate gradient algorithm to solve the optimization
problem with respect to $\hat{\V{x}}=\M{C}\cdot\V{x}$. Following this we
have considered using our statistically independent modes to solve the
problem with respect to the variables $\V{u}=\M{K}^{-1}\cdot\V{w}$. In this
case, it is however advantageous in terms of the number of floating points
operations to use an \emph{explicit} change of variables and to directly
solve the problem for $\V{u}$ rather than for $\V{w}$ with a preconditioner
$\M{M}=\M{K}\mT\cdot\M{K}^{-1}$. Introducing this change of variable in
\Eq{eq:linear-system} and using \Eq{eq:regul_with_u}, the system to solve
becomes:
\begin{equation}
  \left(\M{K}\T\cdot\M{S}\T\cdot\Cerror^{-1}\cdot\M{S}\cdot\M{K}
         +\M{I}\right)\cdot\V{u}
  = \left(\M{K}\T\cdot\M{S}\T\cdot\Cerror^{-1}\cdot\V{d}\right).
\end{equation}
After $\V{u}$ is found by the iterative algorithm, the restored wavefront
is given $\V{w}=\M{K}\cdot\V{u}$.  We expect improvements in the
convergence of the iterative method by using $\V{u}$ instead of $\V{w}$
because this yields an \emph{a priori} covariance matrix equal to the
identity matrix \cite{Skilling_Bryan-1984-maximum_entropy}.  Improved
speedup may be still possible by using a preconditioner on $\V{u}$ as we
discuss in the following.

%
%

\subsection{Diagonal preconditioners}

Diagonal preconditioners may not be the most efficient ones but are very
cheap to use \cite{templates} and are thus considered here.  When the
variable $\V{x}$ in \Eq{eq:Ax=b} follows known statistics, an optimal
preconditioner $\M{M}$ can be computed so that $\M{M}^{-1}\cdot\M{A}$ is,
on average, as close as possible to the identity matrix.  This
\emph{closeness} can be measured in two different spaces: in the data space
or in the parameter space.

In the \emph{data space}, this criterion is written:
\begin{align}
  & \M{M} = \argmin_{\M{M}}\avg{\norm{\M{A}\cdot\V{x}-\M{M}\cdot\V{x}}^{2}}
  \nonumber \\
  \Longleftrightarrow\quad
  & 0 = \frac{\partial\avg{\norm{(\M{A} - \M{M})\cdot\V{x}}^{2}}}
             {\partial\M{M}}
      = 2\,(\M{M} - \M{A})\cdot\avg{\V{x}\cdot\V{x}\T}
  \nonumber \\
  \Longleftrightarrow\quad
  &\M{M}\cdot\M{C}_{\V{x}} = \M{A}\cdot\M{C}_{\V{x}} \, ,
  \label{eq:data-space-precond}
\end{align}
where $\M{C}_{\V{x}}\bydef\avg{\V{x}\cdot\V{x}\T}$ is the covariance matrix
of $\V{x}$.  Of course, if $\M{M}$ is allowed to be any matrix and since
$\M{C}_{\V{x}}$ has full rank, the solution to \Eq{eq:data-space-precond}
is $\M{M}=\M{A}$.  However, for a diagonal preconditioner,
$\M{M}=\diag(\V{m})$, only the diagonal terms of \Eq{eq:data-space-precond}
have to be considered; this yields:
\begin{equation}
  \M{M} = \diag(\V{m})
        = \diag(\M{A}\cdot\M{C}_{\V{x}})\cdot\diag(\M{C}_{\V{x}})^{-1} \,.
\end{equation}
For $\V{x}=\V{u}\sim\NormalLaw\!(\V{0},\M{I})$ then $\M{C}_{\V{x}}=\M{I}$
and \Eq{eq:data-space-precond} simplifies to:
\begin{equation}
  \label{eq:Jacobi-precond}
  \M{M} = \diag(\M{A}) \, ,
\end{equation}
which is the well known Jacobi preconditioner \cite{templates}.

Taking $\M{Q}\bydef\M{M}^{-1}$ and minimizing the statistical distance in
the \emph{parameter space} yields:
\begin{align}
  & \M{Q} = \argmin_{\M{Q}}
    \avg{\norm{\M{Q}\cdot\M{A}\cdot\V{x} - \V{x}}^{2}}
  \nonumber \\
  \Longleftrightarrow\quad
  & 0 = \frac{\partial\avg{\norm{\M{Q}\cdot\M{A}\cdot\V{x} - \V{x}}^{2}}}
             {\partial\M{Q}}
     = 2\,\left(\M{Q}\cdot\M{A} - \M{I}\right)
       \cdot\M{C}_{\V{x}}\cdot\M{A}\T
  \nonumber \\
  \Longleftrightarrow\quad
  &  \M{Q}\cdot\M{A}\cdot\M{C}_{\V{x}}\cdot\M{A}\T
   = \M{C}_{\V{x}}\cdot\M{A}\T \, .
  \label{eq:param-space-precond}
\end{align}
For a diagonal preconditioner, $\M{Q}=\diag(\V{q})$, only the diagonal
terms of \Eq{eq:param-space-precond} have to be considered; hence:
\begin{equation}
  \M{Q} = \diag(\V{q}) = \diag(\M{C}_{\V{x}}\cdot\M{A}\T)
  \cdot\diag(\M{A}\cdot\M{C}_{\V{x}}\cdot\M{A}\T)^{-1} \, .
\end{equation}
Finally, when $\V{x}=\V{u}\sim\NormalLaw\!(\V{0},\M{I})$:
\begin{equation}
  \label{eq:optimal-precond}
  Q_{i,i} =  \frac{A_{i,i}}{\sum_{j}A_{i,j}^{2}}, \qquad\text{and}\qquad
  Q_{i,j\neq i} = 0.
\end{equation}
In contrast to the Jacobi preconditioner, the \emph{optimal} preconditioner
$\M{Q}$ is expensive to compute since every element of matrix $\M{A}$ must
be evaluated to evaluate the denominator.  This however has to be done only
once and for all for a given left-hand side matrix $\M{A}$.  The improvements
given by the diagonal preconditioners in \Eq{eq:Jacobi-precond} and
\Eq{eq:optimal-precond} are compared in the next section.

\section{Simulations and Results}
\label{sec:results}

\subsection{Summary of the various possibilities}

Our previous study gives rise to 6 different possibilities to solve
\Eq{eq:Ax=b}. The first method is based on \Eq{eq:regul_with_u} to
iteratively solve for $\V{w}$:
\begin{equation}
  \label{eq:CG_for_w}
  \left(\M{S}\T\cdot\Cerror^{-1}\cdot\M{S}
         +\M{K}\mT\cdot\M{K}^{-1}\right)\cdot\V{w}
  = \M{S}\T\cdot\Cerror^{-1}\cdot\V{d} \, ,
\end{equation}
using the sparse model matrix $\M{S}$ and the fractal operators
$\M{K}^{-1}$ and $\M{K}\mT$ introduced in
Sect.~\ref{sec:sensor-model} and Sect.~\ref{sec:fractal_operators}.
Although the a priori covariance matrix of $\V{w}$ is not the identity, we
have tried two other methods by assessing the speedup brought by each of
the two diagonal preconditioners defined in \Eq{eq:Jacobi-precond} and
\Eq{eq:optimal-precond}, with
$\M{A}=\M{S}\T\cdot\Cerror^{-1}\cdot\M{S}+\M{K}\mT\cdot\M{K}^{-1}$.

Solving the problem in our statistically independent modes corresponds to a
forth method, requiring to iteratively solve:
\begin{equation}
  \label{eq:CG_for_u}
  \left(\M{K}\T\cdot\M{S}\T\cdot\Cerror^{-1}\cdot\M{S}\cdot\M{K}
         +\M{I}\right)\cdot\V{u}
  = \M{K}\T\cdot\M{S}\T\cdot\Cerror^{-1}\cdot\V{d}
\end{equation}
for $\V{u}$ and then do $\V{w}=\M{K}\cdot\V{u}$. For the two last methods,
we use with \Eq{eq:CG_for_u}, one of the two preconditioners defined in
\Eq{eq:Jacobi-precond} and \Eq{eq:optimal-precond} with
$\M{A}=\M{K}\T\cdot\M{S}\T\cdot\Cerror^{-1}\cdot\M{S}\cdot\M{K}+\M{I}$. In
this case, $\M{C}_{\V{u}}=\M{I}$ so we expect somewhat faster convergence.

\subsection{Comparison of the rates of convergence}

When comparing the efficiency of the six different possibilities, we need
to take into account that the number of floating point operations may be
different for each of them. The aim is not to derive an accurate number of
operations which would depend on the specific implementation of the
algorithms, but rather to get a general estimate. For instance, the
dependence of the $\M{K}$ on $r_{0}$ can be factorized out and included in
operator $\Cerror$ with no extra computational cost. This kind of
optimization was not considered here. As detailed in
Appendix~\ref{sec:computational-burden}, the number of operations is
marginally increased by the preconditioning and does not depend on which
variables ($\V{w}$ or $\V{u}$) are used when starting from an arbitrary
initial vector. A small difference only appears when starting the
algorithms with an initial zero vector, as summarized in
Table~\ref{tab:flops}.

For wavefront reconstruction, when comparing the total number of
operations, $\Nops$, for a given number of (P)CG iterations,
$\Niter$, such that $\Niter\ge1$, we will use these equations:
\begin{equation}
  \begin{split}
    \Nops_\RM{CG} & \sim(\Noverhead+33\,\Niter_\RM{CG})\,\Nparam\,,\\
    \Nops_\RM{PCG} & \sim(\Noverhead+34\,\Niter_\RM{PCG})\,\Nparam\,,    
  \end{split}
  \label{eq:nb_of_ops}
\end{equation}
where $\Noverhead = 4$ when working with variable $\V{w}$, and
$\Noverhead = 10$ when explicitly working with variable $\V{u}$.

In order to assess the speed of the reconstruction, we have tested the
different wavefront reconstruction algorithms on a number of different
conditions. For every simulation, the wavefront sensor sampling is such
that the size of the Shack-Hartmann subaperture is equal to Fried parameter
$r_{0}$. A wavefront is first generated by applying the fractal operator
$\M{K}$ to a vector of normally distributed random values like in
section~\ref{sec:generation_smaller_scales}. The measurements are then
estimated using the current wavefront sensor model, $\M{S}$, and a
stationary uncorrelated random noise $\V{n}$ is added to the simulated
slopes in accordance with \Eq{eq:data-model}. The noise level is given by
its standard deviation $\sigma_{\mathrm{noise}}$ in radians per
subaperture, where the radians here correspond to phase differences between
the edges of the subapertures. At each iteration of the algorithm, the
residual wavefront is computed as the difference between the current
solution and the initial wavefront. The root mean squared error of the
residual wavefront is computed over the pupil, piston removed. The piston
mode is the only removed mode. A central obscuration is always introduced,
with a diameter $1/3$ the diameter of the pupil.

\begin{figure}
  \centerline{
    \includegraphics[width=80mm,keepaspectratio]%
    {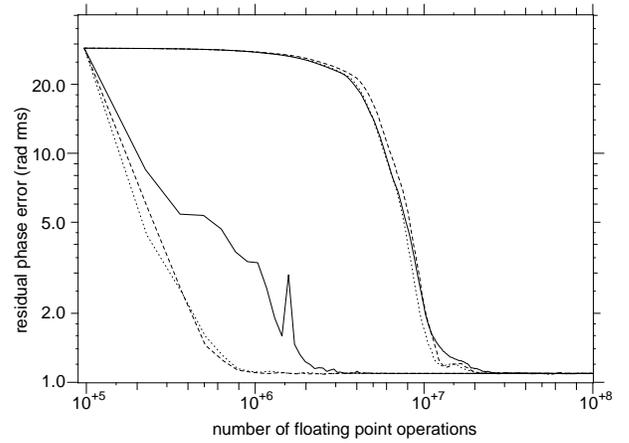} }
    \caption{\label{fig:phase-error-65-stdev1} Phase error as a function of
    the number of operations. Curves are the median value of 100
    simulations with $D/r_{0}=65$, $\sigma_{\mathrm{noise}}=1$
    rad/subaperture and $r_{0}$ has the same size as one subaperture. Solid
    curves are for CG, dashed curves are for PCG with Jacobi
    preconditioner, dotted curves are for PCG with optimal diagonal
    preconditioner. Thin curves are for (P)CG onto the wavefront samples
    $\V{w}$, whereas thick curves are for (P)CG onto the wavefront
    generator $\V{u}$.}
\end{figure}

\begin{figure}
  \centerline{
    \includegraphics[width=80mm,keepaspectratio]%
    {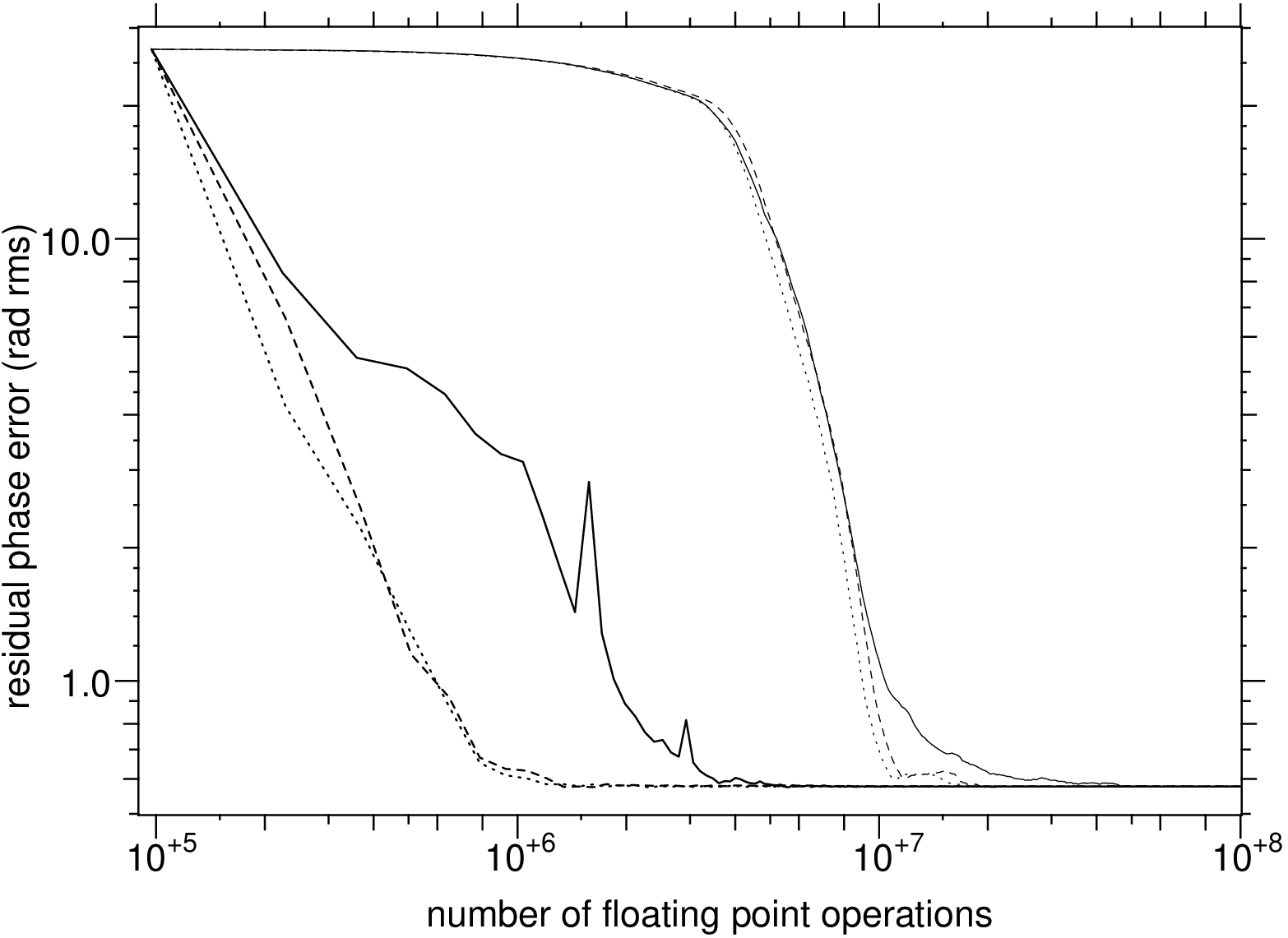} }
    \caption{\label{fig:phase-error-65-stdev0.5} Same as
    Fig.~\ref{fig:phase-error-65-stdev1} but for
    $\sigma_{\mathrm{noise}}=0.5$ rad/subaperture.}
\end{figure}

\begin{figure}
  \centerline{
    \includegraphics[width=80mm,keepaspectratio]%
		    {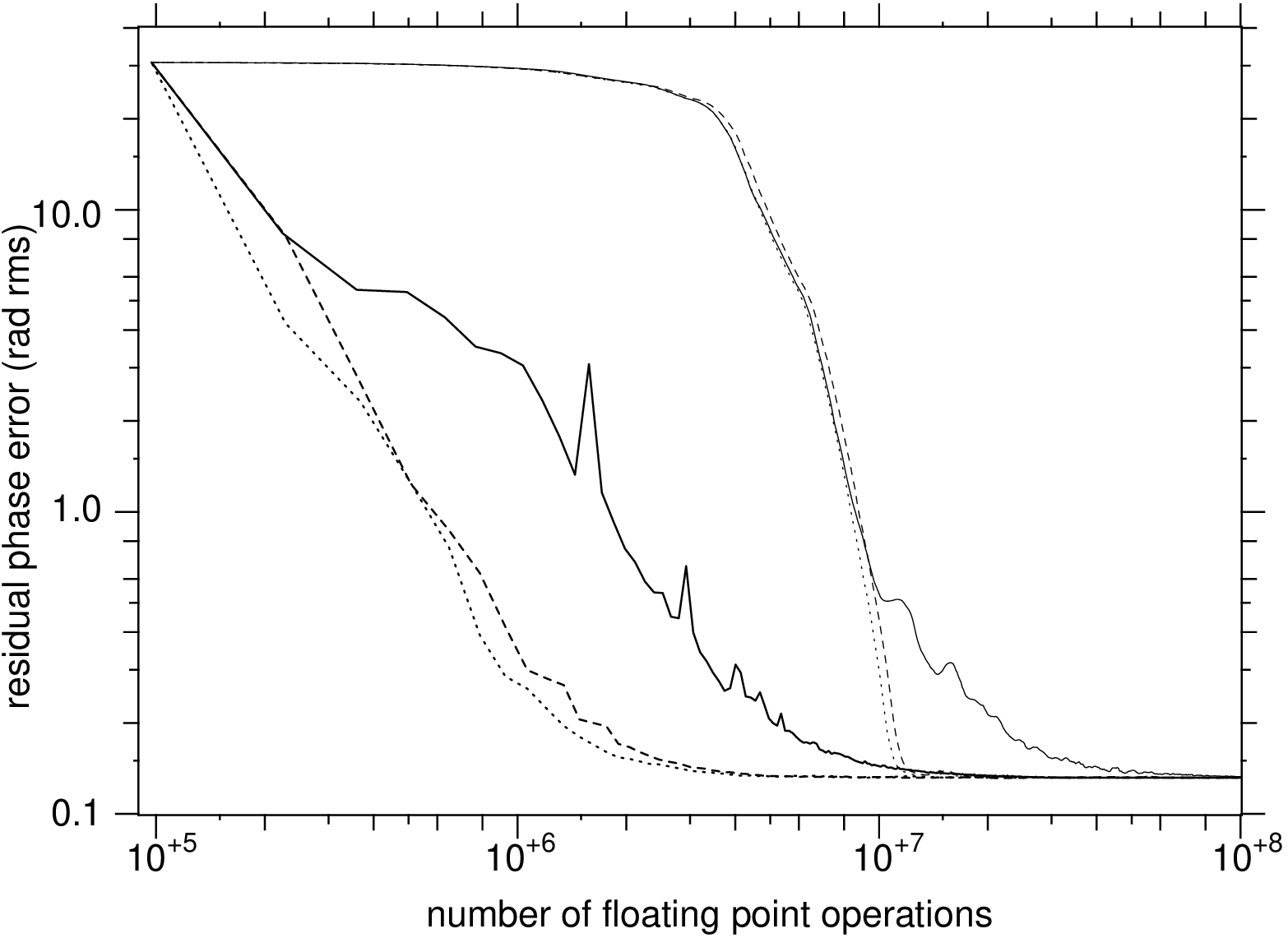}
  }
  \caption{\label{fig:phase-error-65-stdev0.1} Same as
    Fig.~\ref{fig:phase-error-65-stdev1} but for
    $\sigma_{\mathrm{noise}}=0.1$ rad/subaperture.}
\end{figure}

\begin{figure}
  \centerline{
    \includegraphics[width=80mm,keepaspectratio]%
		    {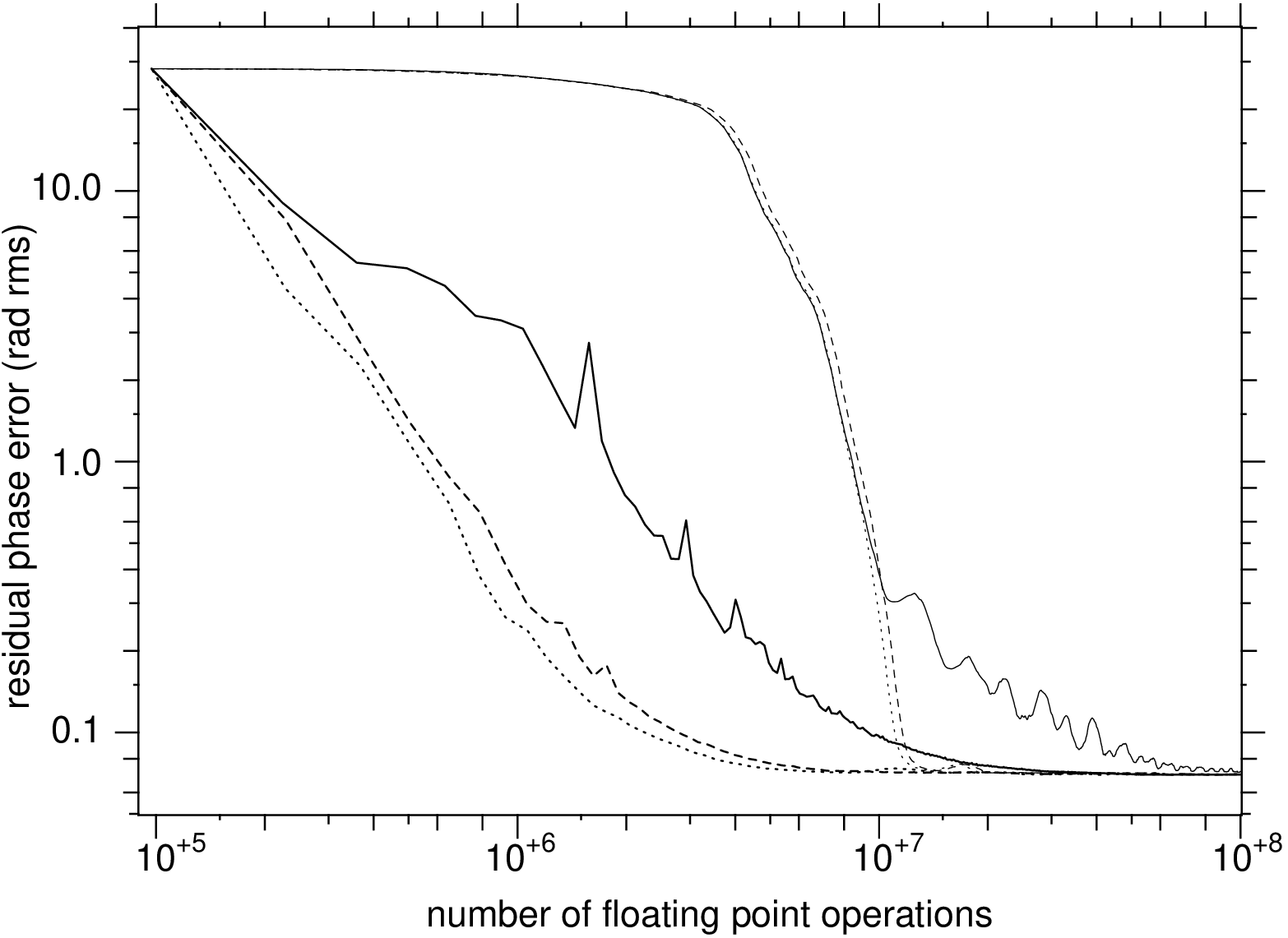}
  }
  \caption{\label{fig:phase-error-65-stdev0.05} Same as
    Fig.~\ref{fig:phase-error-65-stdev1} but for
    $\sigma_{\mathrm{noise}}=0.05$ rad/subaperture.}
\end{figure}

\begin{figure}
  \centerline{
    \includegraphics[width=80mm,keepaspectratio]%
		    {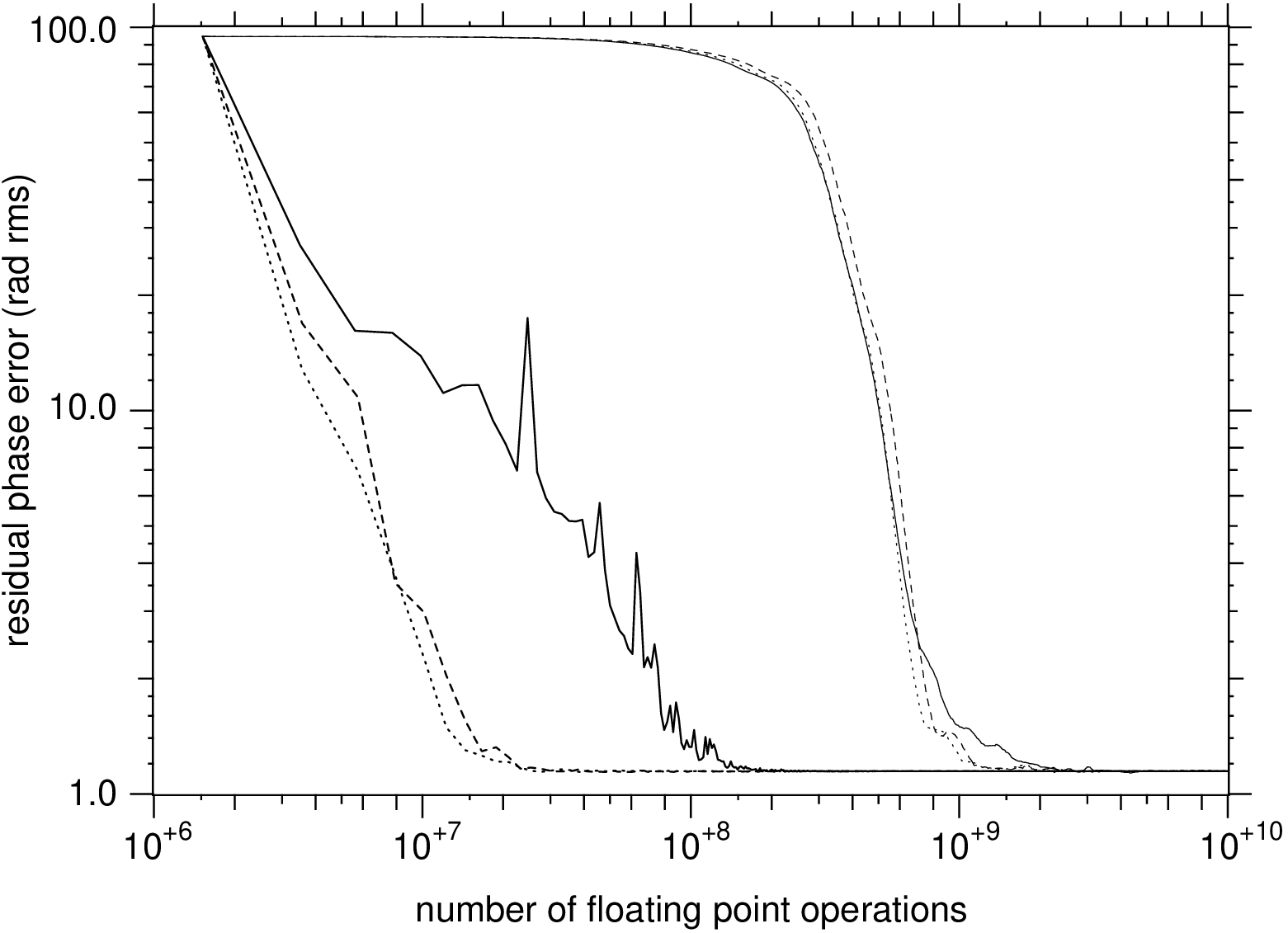}
  }
  \caption{\label{fig:phase-error-257-stdev1} Same as
    Fig.~\ref{fig:phase-error-65-stdev1} but for $D/r_{0}=257$ and
    $\sigma_{\mathrm{noise}}=1$ rad/subaperture.}
\end{figure}

\begin{figure}
  \centerline{
    \includegraphics[width=80mm,keepaspectratio]%
		    {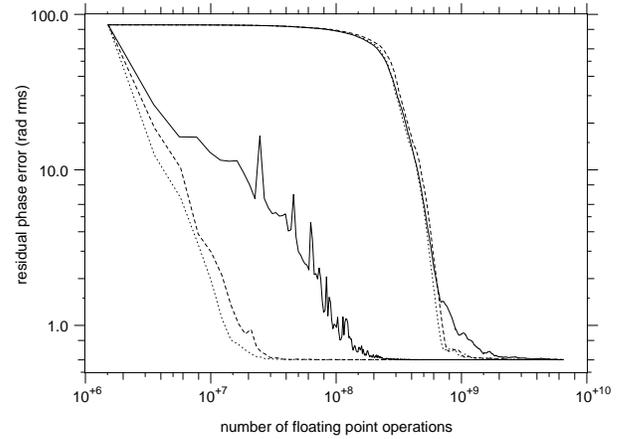}
  }
  \caption{\label{fig:phase-error-257-stdev0.5} Same as
    Fig.~\ref{fig:phase-error-65-stdev1} but for $D/r_{0}=257$ and
    $\sigma_{\mathrm{noise}}=0.5$ rad/subaperture.}
\end{figure}

The graphs presented are for two AO system of size $65\times{}65$ (\cf
Figures \ref{fig:phase-error-65-stdev1}, \ref{fig:phase-error-65-stdev0.5},
\ref{fig:phase-error-65-stdev0.1}, and \ref{fig:phase-error-65-stdev0.05})
and $257\times{}257$ (\cf Figures \ref{fig:phase-error-257-stdev1}, and
\ref{fig:phase-error-257-stdev0.5}). Several levels of noise from
1\,rad/subaperture down to 0.05\,rad/subaperture are examined. They
correspond to the levels of photon noise obtained with $\sim7$ to
$\sim3000$ detected photons per subaperture.
%
%
%
%
%
%
%
%
%
%
On each curve, the 6 algorithms are compared. All the curves plot the
median value obtained for 100 simulations under the same conditions. The
different algorithms were applied to the same simulated wavefronts and
sensor data. The graphs have been plotted assuming a number of floating
point operations given by \Eq{eq:nb_of_ops}, where here the number of
unknowns is $\Nparam = 4225$ and $\Nparam = 66049$ for AO systems
$65\times{}65$ and $257\times{}257$ respectively. Various observations can
be drawn from these curves as dicussed in what follows.

Solving by using $\V{w}$ as unknowns is much slower than using $\V{u}$, by
more than one order of magnitude for a $65\times{}65$ system, and 2 orders
of magnitude for $257\times{}257$. This demonstrates a stunning efficiency
for the fractal operator used as a preconditioner. With $\V{w}$, the
algorithm does not show any improvement of the residual error for a long
time before finding its way toward the solution. In contrast, the very
first steps with $\V{u}$ already show a tremendous reduction of the
residual error. For instance, this feature is critical if the number of
iterations is to be limited to a fixed value as could be the case in
closed-loop.


Using Jacobi or optimal diagonal preconditioners has not the same effect
when working in $\V{w}$ or in $\V{u}$ space. When solving for $\V{w}$, the
preconditioners are only useful at the very end of the convergence, mainly
in the case of high signal-to-noise ratio. Thus they are not very helpful to
reduce the computational load. In contrast, the effect of the diagonal
preconditioners is very effective from the beginning when working with
$\V{u}$. We may notice that the difference between the two diagonal
preconditioners is significant but not critical. The optimal diagonal
preconditioner yields slightly faster convergence.

When $\sigma_{\mathrm{noise}}$ decreases, the convergence of the two
fastest methods takes longer to reach a lower level of residual errors but
the rate of convergence keeps steady. This is analyzed in more detail in
the next section.

\subsection{Number of iterations}

%
%
%
%
%
%
\begin{figure}
  \centering
  \includegraphics[width=80mm,keepaspectratio]%
                  {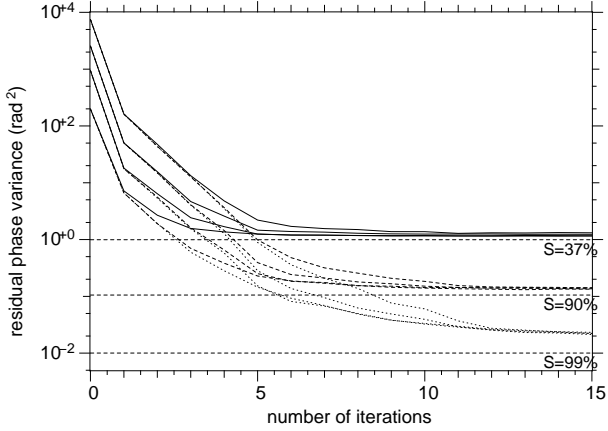}
  \caption{\label{fig:number_iter_method_5} Decrease of the residual phase
    variance as a function of the number of iterations when using $\V{u}$
    as unknowns and optimal diagonal preconditioner. Each curve is the
    median value of 100 simulations.  Three sets of curves are plotted
    for different values of $\sigma^2_{\mathrm{noise}}$: $1$ (solid),
    $0.09$ (dashed), and $0.01\,\mathrm{rad}^2/r_0$ (dotted).  In each set
    of curves, the size of the system increases from bottom to top: 32, 64,
    128 and 256 subapertures along the diameter of the pupil.  Levels of
    Strehl ratios are indicated. The curves show that 5 to 10 iterations
    are enough in most cases for a full reconstruction.}
\end{figure}

%
%
%
%
%
%
%
%
%
%
%
%
\begin{figure}[t]
  \centering
  \includegraphics[width=80mm,keepaspectratio]%
                  {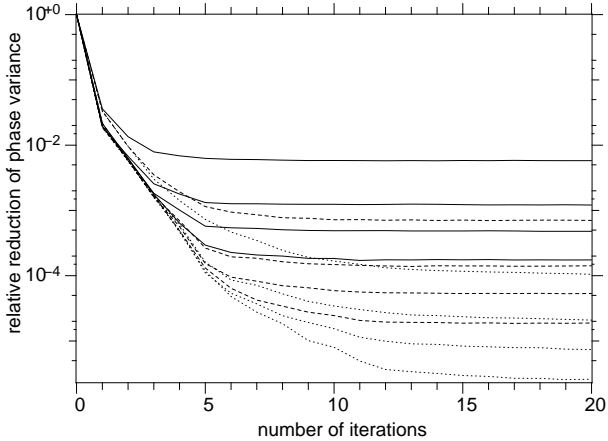}
  \caption{\label{fig:number_iter_method_5_norm}The same curves as those
    in \Fig{fig:number_iter_method_5} are plotted here, normalized by the
    initial variance of the phase. This shows a high relative attenuation
    ($\sim 1/40$) after the first iteration, in any configuration. In each
    set of curves: $\sigma^2_{\mathrm{noise}}=1$ (solid), $0.09$
      (dashed), and $0.01\,\mathrm{rad}^2/r_0$ (dotted); the size of the
    system increases from top to bottom: 32, 64, 128, and 256 subapertures
    along the diameter of the pupil.}
\end{figure}

From the previous section, we now consider only the fastest method, using
both $\V{u}$ as unknowns and the optimal diagonal preconditioner. The aim
here is to assess the number of iterations needed to restore the wavefront.
As in the previous section, we consider one subaperture per $r_{0}$, so the
variance of the incoming wavefronts increases with the size of the system.
Figure \ref{fig:number_iter_method_5} shows how the residual phase variance
decreases at each iteration for various configurations of the system, in
size ($33\times{}33$, $65\times{}65$, $129\times{}129$, $257\times{}257$),
and in noise level ($\sigma^2_{\mathrm{noise}}=1$, $0.09$ and
$0.01\,\mathrm{rad}^2/r_0$). In the first iterations, we can see that the
behavior of the algorithm does not depend on the signal to noise ratio. In
contrast, the final value obtained does not depend on the size of the
system. Strehl levels corresponding to the residual phase variance are
indicated. The curves show that, whatever the size of the system, only 5 to
10 iterations are enough for a reconstruction starting from zero.

%
%
%

In order to remove the effect of starting from different initial phase
variances, the same curves have been normalized by the initial variance on
\Fig{fig:number_iter_method_5_norm}. We can see that the descent of \FRIM
follows the same path for all the simulations, and is stopped at different
values of the final variance, which depends on the signal to noise ratio.
Along this path, the variance is already reduced by a factor $\sim 1/50$ at
the first iteration, $\sim 1/170$ at the second iteration and more than
$\sim 10^{-4}$ at iteration 6. This steep descent will be an asset in
closed-loop.

\section{Conclusion}

We have introduced \FRIM, a new minimum variance iterative algorithm for
fast wavefront reconstruction and fast control of an adaptive optics
system.  Combining fast regularization and efficient preconditioning,
regularized wavefront reconstruction by \FRIM is an $\ScaleAs(N)$
process, where $N$ is the number of wavefront samples.

\FRIM takes advantage of the sparsity of the model matrix $\M{S}$ of
wavefront sensors (or interaction matrices) and makes use of a "fractal
operator" $\M{K}$ for fast computation of the priors.  Based on a
generalization of the mid-point algorithm \cite{LaneEtAl:1992}, $\M{K}$ is
not sparse but is implemented so that it requires only
$\ScaleAs(\Nparam)\simeq 6\,\Nparam$ operations.  Our modifications with
respect to the original algorithm allow the operator to be invertible and
the generated wavefront to be stationary.  We have derived algorithms for
computing $\M{K}^{-1}$, $\M{K}\T$ and $\M{K}\mT$ in the same
number of floating point operations.  In our simulations, we consider a
modified Kolmogorov law but any stationary structure function or covariance
can be implemented in our approach.  The property of stationarity is
expected to be helpful for turbulence tomography.

Another breakthrough comes from the efficiency of the fractal operator when
used as a preconditioner. Combining a fractal change of variables and an
optimal diagonal preconditioner, we were able to reduce the number of
iterations in the range of 5 -- 10 for a full wavefront reconstruction
whatever is the size of the AO system. The exact number of iterations
mainly depends on the signal to noise ratio of the measurements.

It is beyond this work to compare with all the other methods currently
studied in response to the huge increasing of the number of degrees of
freedom for the AO system on ELTs.  Nevertheless, we can easily compare to
standard vector matrix multiplication (VMM).  Assuming uncorrelated noise,
the simulations show that the number of operations with \FRIM is
$\Nops\sim(23+34\,\Niter)\,\Nparam$, where the number of PCG iterations is
$\Niter\lesssim 10$ for any number of degrees of freedom $\Nparam$.
For up to $N=1.3\times10^{4}$ degrees of freedom (\ie $D/r_{0}\le128$), one
wavefront estimation (from scratch) involves $\lesssim 6\times 10^{6}$
operations, that is a bandwidth of $\sim 500\,\mathrm{Hz}$ for a machine
capable of $3\,\mathrm{Gflops}$ which is typical of current workstations.
Conversely, conventional (non-sparse) matrix multiplication would require
$\sim 4\,N^{2}\sim 7\times 10^{8}$ operations to compute the wavefront: our
method is more than 100 times faster.  Furthermore, since the operations
can be done \emph{in-place}, it is expected that the computation with \FRIM
could all be done in cache memory.


For simulating very large AO systems (\eg atmospheric tomography on ELT's),
the speed of the current version of \FRIM is already an asset.  For
real-time control of AO systems, \FRIM algorithm can be parallelized to run
on several CPU's. Being an iterative method (unlike Fourier methods), \FRIM
could be used to improve the estimation of the wavefront from any pieces of
new data as soon as it becomes available.  Hence, \FRIM does not need all
the measurements in a closed-loop system.  A fast iterative method that
gives intermediate results with only a part of the measurements opens the
way to new control approaches for reducing the effect of the delay.  A
further advantage of \FRIM is that it accounts for the statistics of the
turbulence which not only yields a better estimation of the residual phase
\cite{Bechet_et_al-2009-MAP} but also helps to disentangle ambiguities such
as unseen modes in atmospheric tomography.  In this paper, we assume that
the structure function is perfectly known.  Bechet \cite{Bechet-2008-these}
has shown that it is sufficient to not overestimate $r_0$ by more than a
factor $\sim2$ to benefit from the advantages of taking into the priors.

The next step of this work is to extend the theory to closed-loop and to
assess the performances and the properties of the algorithm in this regime.
Since the wavefront is not allowed to change a lot from one step of the AO
loop to the other, the algorithm will always starts close to the solution:
the number of iterations is expected to be yet lower.  This study is not
yet completed but preliminary results have proved the efficiency of \FRIM
in the case of closed-loop adaptive optics
\cite{Bechet_et_al-2006-FRIM_closed_loop,
  Bechet_et_al-2007-FRIM_closed_loop}.


\begin{acknowledgments}
This project forms part of the "ELT Design Study" and is supported by the
European Commission, within its Framework Programme 6, under contract No
011863. This work was also supported by contract No 0712729 with ESO, The
European Southern Observatory.

The authors would like to thank Cl\'ementine B\'echet and Nicholas Devaney
for their fruitful comments.
The algorithms and the simulations presented in this article have been
implemented in Yorick, a free data processing language written by David
Munro (\href{http://yorick.sourceforge.net/}{http://yorick.sourceforge.net/}).
\end{acknowledgments}


\appendix

\section{Derivation of the interpolation coefficients}
\label{sec:coefs}

In this appendix, we detail the computation of the interpolation
coefficients involved in the different configurations shown by
\Fig{fig:fractal-step1}. Denoting $r$ the step size in the grid before the
refinement, the distances between the points considered in this refinement
step are: $\SqrtTwo\,r$, $r$, $r/\SqrtTwo$, or $r/2$
(\Fig{fig:fractal-step1}). Hence the only covariances required in our
computations are:
\begin{equation}
  \begin{array}{rclcl}
    c_0 &=& c(0)           &=& \sigma^{2} \\
    c_1 &=& c(r/2)         &=& \sigma^{2} - f(r/2)/2 \\
    c_2 &=& c(r/\SqrtTwo)  &=& \sigma^{2} - f(r/\SqrtTwo)/2 \\
    c_3 &=& c(r)           &=& \sigma^{2} - f(r)/2 \\
    c_4 &=& c(\SqrtTwo\,r) &=& \sigma^{2} - f(\SqrtTwo\,r)/2 \\
  \end{array}
\end{equation}
where $c(r)$ and $f(r)$ are respectively the covariance and the structure
function for a separation $r$.

\subsection{Square configuration}

For the interpolation stage illustrated by the top-left part of
\Fig{fig:fractal-step1} and according to \Eq{eq:alpha-system-covar}, the
interpolation coefficients $\{\alpha_1,\alpha_2,\alpha_3,\alpha_4\}$ are
obtained by solving:
\begin{displaymath}
  \Matrix{cccc}{
    c_0 & c_3 & c_4 & c_3 \\
    c_3 & c_0 & c_3 & c_4 \\
    c_4 & c_3 & c_0 & c_3 \\
    c_3 & c_4 & c_3 & c_0 \\
  }\cdot\Matrix{c}{
    \alpha_1 \\
    \alpha_2 \\
    \alpha_3 \\
    \alpha_4 \\
  } = \Matrix{c}{
    c_2 \\
    c_2 \\
    c_2 \\
    c_2 \\
  }\,.
\end{displaymath}
Solving this linear system and plugging the solution into
\Eq{eq:alpha-system-covar} leads to:
\begin{equation}
  \label{eq:coef-square-config}
  \renewcommand{\TmpI}{c_0}
  \renewcommand{\TmpII}{c_2}
  \renewcommand{\TmpIII}{c_3}
  \renewcommand{\TmpIV}{c_4}
  \renewcommand{\TmpV}{\TmpI + 2\,\TmpIII + \TmpIV}
  \begin{split}
    \alpha_1 = \alpha_2 = \alpha_3 = \alpha_4 &= \frac{\TmpII}{\TmpV}\,, \\
    \alpha_0 &= \pm\sqrt{\TmpI - \frac{4\,\TmpII^2}{\TmpV}}\,. \\
  \end{split}
\end{equation}
Note that the sign of $\alpha_0$ is irrelevant.

\subsection{Triangle configuration}

In original mid-point algorithm \cite{LaneEtAl:1992}, the values at the
edges of the support (top-right part of \Fig{fig:fractal-step1}) were
generated from only the two neighbors on the edge, ignoring the third
interior neighbor (denoted $w_3$ in the figure). Here, according to
\Eq{eq:alpha-system-covar}, the interpolation coefficients
$\{\alpha_1,\alpha_2,\alpha_3\}$ for this stage are obtained by solving:
\begin{displaymath}
  \Matrix{ccc}{
    c_0 & c_3 & c_2 \\
    c_3 & c_0 & c_2 \\
    c_2 & c_2 & c_0 \\
  }\cdot\Matrix{c}{
    \alpha_1 \\
    \alpha_2 \\
    \alpha_3 \\
  } = \Matrix{c}{
    c_1 \\
    c_1 \\
    c_1 \\
  }\,.
\end{displaymath}
Solving this linear system and plugging the solution into
\Eq{eq:alpha-system-covar} leads to:
\begin{equation}
  \label{eq:coef-triangle-config}
  \renewcommand{\TmpI}{c_0\,(c_0 + c_3) - 2\,c_2^2}
  \begin{split}
    \alpha_1 = \alpha_2
    &= \frac{c_1\,(c_0 - c_2)}{\TmpI} \\
    \alpha_3
    &= \frac{c_1\,(c_0 - 2\,c_2 + c_3)}{\TmpI} \\
    \alpha_0
    &= \pm\sqrt{c_0 - \frac{c_1^2\,(3\,c_0 - 4\,c_2 + c_3)}{\TmpI}}\\
  \end{split}
\end{equation}

\subsection{Diamond configuration}

The interpolation coefficients for the stage in the bottom part of
\Fig{fig:fractal-step1} can be deduced from \Eq{eq:coef-square-config} by
replacing $r$ by $r/\SqrtTwo$, then:
\begin{equation}
  \label{eq:coef-diamond-config}
  \renewcommand{\TmpI}{c_0}
  \renewcommand{\TmpII}{c_1}
  \renewcommand{\TmpIII}{c_2}
  \renewcommand{\TmpIV}{c_3}
  \renewcommand{\TmpV}{\TmpI + 2\,\TmpIII + \TmpIV}
  \begin{split}
    \alpha_1 = \alpha_2 = \alpha_3 = \alpha_4 &= \frac{\TmpII}{\TmpV}\,, \\
    \alpha_0 &= \pm\sqrt{\TmpI - \frac{4\,\TmpII^2}{\TmpV}}\,. \\
  \end{split}
\end{equation}
%

\section{Computational Burden}
\label{sec:computational-burden}

In order to estimate the number of floating point operations, we need to
carefully detail the steps of the CG method and count the number of
operations involved in the multiplication by the different linear operators
$\M{S}$, $\M{K}$, \etc. Figure~\ref{fig:conjugate-gradient-algorithm}
summarizes the steps of the (PCG) algorithm \cite{templates} to solve
\Eq{eq:Ax=b}.

\begin{table}[t]
  \begin{tabular}{|rc|}
    \hline
    \multicolumn{1}{|c}{algorithm step} & floating point operations \\
    \hline
    \hline
    initialization:\hfill general case & $\sim 25\,\Nparam$ \\
    zero initial vector in $\V{u}$ space & $\sim 12\,\Nparam$ \\
    zero initial vector in $\V{w}$ space & $\sim 6\,\Nparam$ \\
    \hline
    1st CG iteration &  $\sim 31\,\Nparam$ \\
    any subsequent CG iteration & $\sim 33\,\Nparam$ \\
    total after $\Niter\ge1$ iterations
    & $\sim(23 + 33\,\Niter)\,\Nparam$ \\
    \hline
    1st PCG iteration & $\sim 32\,\Nparam$ \\
    any subsequent PCG iteration & $\sim 34\,\Nparam$ \\
    total after $\Niter\ge1$ iterations
    & $\sim(23 + 34\,\Niter)\,\Nparam$ \\
    \hline
  \end{tabular}
  \caption{\label{tab:flops}
    Number of operations involved in conjugate gradients (CG) and
    preconditioned conjugate gradients (PCG) applied to the wavefront
    restoration problem solved by our algorithm. The integers $\Nparam$
    and $\protect\Niter$ are respectively the number of unknowns and number of
    iterations. For a reconstruction, we assume an initial null guess in
    the initialization step: in this case the number of operations at
    this step is reduced down to $\sim 6\,\Nparam$ or $\sim 12\,\Nparam$
    when respectively $\V{w}$ or $\V{u}$ are used as
    unknowns.}
\end{table}

If the unknowns are the wavefront samples, then $\V{x}=\V{w}$ and:
\begin{align*}
  \M{A} &= \M{S}\T\cdot\Cerror^{-1}\cdot\M{S}
            + \M{K}\mT\cdot\M{K}^{-1} \, , \notag\\
  \V{b} &= \M{S}\T\cdot\Cerror^{-1}\cdot\V{d}    \, . \notag
\end{align*}
Starting the algorithm with $\V{w}_{0}$, the initial residuals write:
\begin{align}
  \V{r}_{0}
  &= \V{b} - \M{A}\cdot\V{w}_{0}\notag\\
  &= \M{S}\T\cdot\Cerror^{-1}\cdot(\V{d} - \M{S}\cdot\V{w}_{0})
  - \M{K}\mT\cdot\M{K}^{-1}\cdot\V{w}_{0}\,.
  \label{eq:r0_w}
\end{align}
%

If the unknowns are the wavefront generators, then $\V{x}=\V{u}$ and:
\begin{align*}
  \M{A} &= \M{K}\T\cdot\M{S}\T\cdot\Cerror^{-1}\cdot\M{S}\cdot\M{K}
           + \M{I}                                        \,, \notag\\
  \V{b} &= \M{K}\T\cdot\M{S}\T\cdot\Cerror^{-1}\cdot\V{d} \,, \notag
\end{align*}
where $\M{I}$ is the identity matrix. Starting the algorithm with
$\V{u}_{0}$, the initial residuals are:
\begin{align}
  \V{r}_{0}
  &= \V{b} - \M{A}\cdot\V{u}_{0}\notag\\
  &= \M{K}\T\cdot\M{S}\T\cdot\Cerror^{-1}\cdot
     (\V{d} - \M{S}\cdot\M{K}\cdot\V{u}_{0}) - \V{u}_{0}\,.
  \label{eq:r0_u}
\end{align}


Making use of possible factorizations (some of the $\alpha_{i}$'s have the
same values), applying any one of the operators $\M{K}$, $\M{K}\T$,
$\M{K}^{-1}$, or $\M{K}\mT$ involves the same number of floating
point operations:
\begin{align*}
  \Nops(\M{K}) &=
  \Nops(\M{K}\T) =
  \Nops(\M{K}^{-1}) =
  \Nops(\M{K}\mT) \\
  & = 6\,\Nparam_{\V{u}} - 14\\
  & \sim 6\,\Nparam\,,
\end{align*}
where $\Nparam$ is the number of degrees of freedom of the system,
$\Nparam_{\V{u}}\sim \Nparam$ is the number of elements in vector $\V{u}$
and, in our notation, $\Nops(\M{L})$ is the number of floating point
operations required to apply a linear operator $\M{L}$ to a vector.

Since we consider uncorrelated data noise, $\Cerror^{-1}$ is diagonal and:
\begin{displaymath}
  \Nops\left(\Cerror^{-1}\right)=\Ndata\sim2\,\Nparam\,;
\end{displaymath}
however note that these $\sim 2\,\Nparam$ floating point operations per
iteration can be saved for stationary noise (\ie
$\Cerror^{-1}\propto\M{I}$).

%
%
%
%
For Fried model of wavefront sensor and after proper factorization:
\begin{displaymath}
  \Nops(\M{S}) = \Nops(\M{S}\T)\sim 4\,\Nparam\,.
\end{displaymath}
This assumes, in particular, that the data were pre-multiplied by 2 (see
\Eq{eq:Fried-model}).


Finally, whatever the unknown are ($\V{w}$ or $\V{u}$), the total number of
floating point operations required to apply the left hand side matrix
$\M{A}$ to a given vector is:
\begin{align*}
  \Nops(\M{A})
  &\sim 2\,\Nops(\M{K}) + 2\,\Nops(\M{S})
    + \Nops(\Cerror^{-1}) + \Nparam \\
  &\sim 23\,\Nparam\,.
\end{align*}
The last $\Nparam$ comes from the addition of likelihood and regularization
terms.

From equations (\ref{eq:r0_w}) and (\ref{eq:r0_u}), using either $\V{w}$ or
$\V{u}$ as the unknowns, initialization of the CG, \ie computation of the
initial residuals $\V{r}_{0}$, involves
\begin{align*}
  \Nops(\V{r}_{0})
  &\sim 2\,\Nops(\M{K}) + 2\,\Nops(\M{S})
    + \Nops(\Cerror^{-1}) + \Ndata + \Nparam \\
  &\sim 25\,\Nparam
\end{align*}
operations. Note that, if the algorithm is initialized with
$\V{x}_{0}=\V{0}$ (a vector of zeroes), this number of operations is
significantly reduced down to $\sim 6\,\Nparam$ and $\sim 12\,\Nparam$ when
respectively $\V{w}$ and $\V{u}$ are used as unknowns.  Also note that
there may be additional $\sim6\,\Nparam$ operations to compute $\V{w}$ from
$\V{u}$ when necessary.

Whatever are the considered variables, the number of unknowns is $\sim
\Nparam$, hence any dot product in the CG algorithm involves
$2\,\Nparam-1\sim2\,\Nparam$ floating point operations.  The first CG
iteration (\Fig{fig:conjugate-gradient-algorithm}) requires two dot
products ($2\,\Nparam-1\sim 2\,\Nparam$ floating point operations each) to
compute $\rho_k$ and $\alpha_k$, applying $\M{A}$ once and two vector
updates (involving $\sim 2\,\Nparam$ operations each); hence a total of
$\sim 31\,\Nparam$ operations. Any subsequent iteration requires an
additional vector update to compute the conjugate gradient direction; hence
$\sim 33\,\Nparam$ operations. Finally, preconditioning by a diagonal
preconditioner simply adds $\sim \Nparam$ operations per iteration.

The number of floating operations required by the different versions of the
reconstruction algorithm are summarized in table~\ref{tab:flops} and
by~\Eq{eq:nb_of_ops}.  Note that in the general case, the number of
operations does not depend on which variables $\V{w}$ or $\V{u}$ are
used. There is a difference of $\sim 6\,\Nparam$ operations in the
initialization step only when the algorithm is started with a zero initial
vector (see table~\ref{tab:flops}).

\ifthenelse{1=0}{%
 \bibliographystyle{osajnl}
 \bibliography{frim}

\begin{thebibliography}{10}
\newcommand{\enquote}[1]{``#1''}

\bibitem{Roddier_1999}
F.~Roddier, \emph{{Adaptive Optics in Astronomy}} (Cambridge University Press,
  1999).

\bibitem{Gendron_&_Lena_1994}
E.~Gendron and P.~L\'{e}na, \enquote{{Astronomical adaptive optics. I. Modal
  control optimization},} Astron. Astrophys. \textbf{291}, 337--347 (1994).

\bibitem{LeLouarn_et_al_2000b}
M.~Le~Louarn, N.~Hubin, M.~Sarazin, and A.~Tokovinin, \enquote{{New challenges
  for adaptive optics: extremely large telescopes},} Mon. Not. R. Astr. Soc.
  \textbf{317}, 535--544 (2000).

\bibitem{Hubin_et_al_2005}
N.~Hubin, B.~L. Ellerbroek, R.~Arsenault, R.~M. Clare, R.~Dekany, L.~Gilles,
  M.~Kasper, G.~Herriot, M.~Le~Louarn, E.~Marchetti, S.~Oberti, J.~Stoesz,
  J.-P. V\'{e}ran, and C.~V\'{e}rinaud, \enquote{{Adaptive optics for Extremely
  Large Telescopes},} in \enquote{Scientific Requirements for Extremely Large
  Telescopes,} , vol. 232 of \emph{IAU Symposium}, P.~A. Whitelock,
  M.~Dennefeld, and B.~Leibundgut, eds. (Cambridge University Press, 2005),
  vol. 232 of \emph{IAU Symposium}, pp. 60--85.

\bibitem{Poyneer_et_al_2002}
L.~A. Poyneer, D.~T. Gavel, and J.~M. Brase, \enquote{{Fast wave-front
  reconstruction in large adaptive optics systems with use of the Fourier
  transform},} J. Opt. Soc. Am. A \textbf{19}, 2100--2111 (2002).

\bibitem{Poyneer_et_al_2008}
L.~A. Poyneer, D.~Dillon, S.~Thomas, and B.~A. Macintosh, \enquote{{Laboratory
  demonstration of accurate and efficient nanometer-level wavefront control for
  extreme adaptive optics},} Appl. Opt. \textbf{47}, 1317--1326 (2008).

\bibitem{MacMartin_2003}
D.~G. MacMartin, \enquote{{Local, hierarchic, and iterative reconstructors for
  adaptive optics},} J. Opt. Soc. Am. A \textbf{20}, 1084--1093 (2003).

\bibitem{LeRoux_et_al_2004}
B.~Le~Roux, J.-M. Conan, C.~Kulcs\'ar, H.-F. Raynaud, L.~M. Mugnier, and
  T.~Fusco, \enquote{{Optimal control law for classical and multiconjugate
  adaptive optics},} J. Opt. Soc. Am. A \textbf{21}, 1261--1276 (2004).

\bibitem{Ellerbroek_2002}
B.~L. Ellerbroek, \enquote{{Efficient computation of minimum-variance
  wave-front reconstructors with sparse matrix techniques},} J. Opt. Soc. Am. A
  \textbf{19}, 1803--1816 (2002).

\bibitem{Vogel_2004}
C.~R. Vogel, \enquote{{Sparse matrix methods for wavefront reconstruction
  revisited},} in \enquote{Advancements in Adaptive Optics,} , vol. 5490 of
  \emph{SPIE Conference}, D.~Bonaccini, B.~L. Ellerbroek, and R.~Ragazzoni,
  eds. (Society of Photo-Optical Instrumentation Engineers, Bellingham, WA,
  2004), vol. 5490 of \emph{SPIE Conference}, pp. 1327--1335.

\bibitem{Southwell_1980}
W.~H. Southwell, \enquote{{Wave-front estimation from wave-front slope
  measurements},} J. Opt. Soc. Am. \textbf{70}, 998--1006 (1980).

\bibitem{Nocedal_Wright-2006-numerical_optimization}
J.~Nocedal and S.~J. Wright, \emph{Numerical Optimization} (Springer Verlag,
  2006), 2nd ed.

\bibitem{NumericalRecipes}
W.~H. Press, S.~A. Teukolsky, W.~T. Vetterling, and B.~P. Flannery,
  \emph{Numerical Recipes in C} (Cambridge University Press, 1992), 2nd ed.

\bibitem{Wild_et_al_1995}
W.~J. Wild, E.~J. Kibblewhite, and R.~Vuilleumier, \enquote{{Sparse matrix
  wave-front estimators for adaptive-optics systems for large ground-based
  telescopes},} Opt. Lett. \textbf{20}, 955--957 (1995).

\bibitem{templates}
R.~Barrett, M.~Berry, T.~F. Chan, J.~Demmel, J.~Donato, J.~Dongarra,
  V.~Eijkhout, R.~Pozo, C.~Romine, and H.~V. der Vorst, \emph{Templates for the
  Solution of Linear Systems: Building Blocks for Iterative Methods} (SIAM,
  Philadelphia, PA, 1994).

\bibitem{Gilles_et_al_2002c}
L.~Gilles, C.~R. Vogel, and B.~L. Ellerbroek, \enquote{{Multigrid
  preconditioned conjugate-gradient method for large-scale wave-front
  reconstruction},} J. Opt. Soc. Am. A \textbf{19}, 1817--1822 (2002).

\bibitem{Gilles_2003b}
L.~Gilles, \enquote{{Order-N sparse minimum-variance open-loop reconstructor
  for extreme adaptive optics},} Opt. Lett. \textbf{28}, 1927--1929 (2003).

\bibitem{Gilles_et_al_2003}
L.~Gilles, B.~L. Ellerbroek, and C.~R. Vogel, \enquote{{Preconditioned
  conjugate gradient wave-front reconstructors for multiconjugate adaptive
  optics},} Appl. Opt. \textbf{42}, 5233--5250 (2003).

\bibitem{Yang_et_al_2006c}
Q.~Yang, C.~R. Vogel, and B.~L. Ellerbroek, \enquote{{Fourier domain
  preconditioned conjugate gradient algorithm for atmospheric tomography},}
  Appl. Opt. \textbf{45}, 5281--5293 (2006).

\bibitem{Vogel_&_Yang_2006c}
C.~R. Vogel and Q.~Yang, \enquote{{Fast optimal wavefront reconstruction for
  multi-conjugate adaptive optics using the Fourier domain preconditioned
  conjugate gradient algorithm},} Opt. Express \textbf{14}, 7487--7498 (2006).

\bibitem{Gilles_et_al_2007}
L.~Gilles, B.~Ellerbroek, and C.~Vogel, \enquote{{A comparison of Multigrid
  V-cycle versus Fourier Domain Preconditioning for Laser Guide Star
  Atmospheric Tomography},} in \enquote{Signal Recovery and Synthesis,} , B.~L.
  Ellerbroek and J.~C. Christou, eds. (Optical Society of America, Washington,
  USA, 2007), OSA topical meetings, p. paper JTuA1.

\bibitem{Gilles_&_Ellerbroek_2008b}
L.~Gilles and B.~L. Ellerbroek, \enquote{{Split atmospheric tomography using
  laser and natural guide stars},} J. Opt. Soc. Am. A \textbf{25}, 2427--2435
  (2008).

\bibitem{LaneEtAl:1992}
R.~G. Lane, A.~Glindemann, and J.~C. Dainty, \enquote{Simulation of a
  kolmogorov phase screen,} Wave in random media \textbf{2}, 209--224 (1992).

\bibitem{Fried_1977}
D.~L. Fried, \enquote{{Least-squares fitting a wave-front distortion estimate
  to an array of phase-difference measurements},} J. Opt. Soc. Am. \textbf{67},
  370--375 (1977).

\bibitem{Thiebaut:2005:Cargese}
E.~Thi\'ebaut, \enquote{Introduction to image reconstruction and inverse
  problems,} in \enquote{Optics in Astrophysics,} , R.~Foy and F.-C. Foy, eds.
  (Springer, Dordrecht, The Netherlands, 2005), NATO ASI, p. 397.

\bibitem{Herrmann_1992}
J.~Herrmann, \enquote{{Phase variance and Strehl ratio in adaptive optics},} J.
  Opt. Soc. Am. A \textbf{9}, 2258--2259 (1992).

\bibitem{Tarantola-2005-inverse_problem_theory}
A.~Tarantola, \emph{Inverse Problem Theory and Methods for Model Parameter
  Estimation} (SIAM, 2005).

\bibitem{TarantolaValette:1982}
A.~Tarantola and B.~Valette, \enquote{Inverse problems = quest for
  information,} Journal of Geophysics \textbf{50}, 159--170 (1982).

\bibitem{Fried_1965}
D.~L. Fried, \enquote{{Statistics of a geometric representation of wavefront
  distortion},} J. Opt. Soc. Am. \textbf{55}, 1427--1435 (1965).

\bibitem{Rousset_1993}
G.~Rousset, \enquote{{Wavefront sensing},} in \enquote{Adaptive optics for
  astronomy,} , vol. 423 of \emph{Proc. NATO ASI Series C}, D.~M. Alloin and
  J.-M. Mariotti, eds. (Kluwer, Dordrecht, The Netherlands, 1993), vol. 423 of
  \emph{Proc. NATO ASI Series C}, pp. 115--137.

\bibitem{Roddier-1990-wavefront_simulation}
N.~Roddier, \enquote{Atmospheric wavefront simulation using zernike
  polynomials,} Opt. Eng. \textbf{29}, 1174--1180 (1990).

\bibitem{Skilling_Bryan-1984-maximum_entropy}
J.~Skilling and R.~K. Bryan, \enquote{Maximum entropy image reconstruction:
  general algorithm,} Monthlty Notices of the Royal Astronomical Society
  \textbf{211}, 111--124 (1984).

\bibitem{Bechet_et_al-2009-MAP}
C.~B\'echet, M.~Tallon, and E.~Thi\'ebaut, \enquote{Comparison of minimum-norm
  maximum likelihood and maximum a posteriori wavefront reconstructions for
  large adaptive optics systems,} \josaa \textbf{26}, 497--508 (2009).

\bibitem{Bechet-2008-these}
C.~B\'echet, \enquote{Commande optimale rapide pour l'optique adaptative des
  futurs t\'elescopes hectom\'etriques,} Ph.D. thesis, Ecole Centrale de Lyon
  (2008).

\bibitem{Bechet_et_al-2006-FRIM_closed_loop}
C.~{B{\'e}chet}, M.~{Tallon}, and E.~{Thi{\'e}baut}, \enquote{{FRIM:
  minimum-variance reconstructor with a fractal iterative method},} in
  \enquote{Advances in Adaptive Optics II.}, , vol. 6272 of \emph{SPIE
  Conference}, D.~B.~C. B.~L.~Ellerbroek, ed. (2006), vol. 6272 of \emph{SPIE
  Conference}, p. 62722U.

\bibitem{Bechet_et_al-2007-FRIM_closed_loop}
C.~{B{\'e}chet}, M.~{Tallon}, and E.~{Thi{\'e}baut}, \enquote{{Closed-Loop AO
  Performance with FrIM},} in \enquote{Adaptive Optics: Analysis and Methods,}
  (2007), Conference of the Optical Society of America, p. JTuA4.

\end{thebibliography}
}{%

}

\end{document}